\documentclass[conference]{IEEEtran}
\usepackage[utf8]{inputenc}
\usepackage{amsmath}
\usepackage{amsfonts}
\usepackage{amssymb}
\usepackage{color}
\usepackage{graphicx}
\usepackage{stmaryrd}
\usepackage{fontawesome}
\usepackage[table]{xcolor}
\usepackage{arydshln}
\usepackage{svg}
\usepackage{rotating}
\usepackage{fancyvrb}

\usepackage{url}
\usepackage{booktabs}
\usepackage[justification=centering]{caption}
\usepackage{pgfplots}
\usepackage{algorithm}
\usepackage[noend]{algpseudocode}

\pgfplotsset{%
    ,compat=1.12
    ,every axis x label/.style={at={(current axis.right of origin)},anchor=north west}
    ,every axis y label/.style={at={(current axis.above origin)},anchor=north east}
    }

\definecolor{pblue}{rgb}{0.13,0.13,1}
\definecolor{pgreen}{rgb}{0,0.5,0}
\definecolor{pred}{rgb}{0.9,0,0}
\definecolor{pgrey}{rgb}{0.46,0.45,0.48}

\definecolor{Gray}{gray}{0.85}

\usepackage{stackengine}
\setstackgap{L}{.5\baselineskip}
\newcommand\markabove[2]{{\sffamily\color{red}\hsmash{$\uparrow$}%
  \smash{\toplap{#1}{\scriptsize\bfseries#2}}}}

\usepackage{ifthen}
\newcounter{todoindex}
\newcommand\TODO[1]{%
  \addtocounter{todoindex}{1}%
  \expandafter\def\csname todo\roman{todoindex}\endcsname{#1}%
  \markabove{c}{\Alph{todoindex}}%
}
\newcounter{index}
\newcommand\showTODOs{%
  \vspace{5ex}%
  \rule{10ex}{.5ex}\textcolor{red}{TO-DO LIST}\rule{10ex}{.5ex}\\%
  \setcounter{index}{0}%
  \whiledo{\value{index} < \value{todoindex}}{%
    \addtocounter{index}{1}%
    \markabove{c}{\Alph{index}}  \csname todo\roman{index}\endcsname\\%
  }%
}

\usepackage{pgf}
\usepackage{tikz}
\usetikzlibrary{arrows.meta,automata, positioning, spy, chains}

\usepackage{listings}
\usepackage{textcomp}
\newcommand{\orsyn}{\ensuremath{\:\mid\:}}
\newcommand{\toolname}[1]{Beagle}

\newcommand{\drag}[1]{\faArrows}
\newcommand{\click}[1]{\faHandPointerO}
\newcommand{\wait}[1]{\faClockO}
\newcommand{\type}[1]{\faKeyboardO}

\newcommand{\casestudyapp}[1]{{\sc scw}}

\newcommand{\qed}{\hfill\ensuremath{\blacksquare}}

\newcounter{example}
\newenvironment{example}[1][]{\refstepcounter{example}\noindent
   \textbf{Example~\theexample. #1} \rmfamily}{\qed}

\newcounter{definition}
\newenvironment{definition}[1][]{\refstepcounter{definition}\noindent
   \textbf{Definition~\thedefinition. #1}}{\qed}

\VerbatimFootnotes

\author{\IEEEauthorblockN{Gabriele Costa}
\IEEEauthorblockA{IMT School for Advanced Studies\\
Lucca, Italy\\
gabriele.costa@imtlucca.it}
\and
\IEEEauthorblockN{Andrea Valenza}
\IEEEauthorblockA{IMQ Minded Security\\
Milan, Italy \\ 
andrea.valenza@mindedsecurity.com}
}

\title{Why Charles Can Pen-test: an Evolutionary Approach to Vulnerability Testing}

\begin{document}

\maketitle

\begin{abstract}
Discovering vulnerabilities in applications of real-world complexity is a daunting task:
a vulnerability may affect a single line of code, and yet it compromises the security of
the entire application. Even worse, vulnerabilities may manifest only in exceptional
circumstances that do not occur in the normal operation of the application. It is widely
recognized that state-of-the-art penetration testing tools play a crucial role, and are
routinely used, to dig up vulnerabilities. Yet penetration testing is still primarily a
human-driven activity, and its effectiveness still depends on the skills and ingenuity of
the security analyst driving the tool.
In this paper, we propose a technique for the automatic discovery of vulnerabilities in
event-based systems, such as web and mobile applications.
Our approach is based on a \emph{collaborative}, \emph{co-evolutionary} and
\emph{contract-driven} search strategy that iteratively $(i)$ executes a pool of
test cases, $(ii)$ identifies the most \emph{promising} ones and $(iii)$ generates new
test cases from them.
The approach makes a synergistic combination of evolutionary algorithms where several
``species'' contribute to solving the problem: one species, the \emph{test species},
evolves to find the target test case, i.e.,~the set of instruction whose execution lead to
the vulnerable statement, whereas the other species, called \emph{contract species},
evolve to select the parameters for the procedure calls needed to trigger the
vulnerability.
To assess the effectiveness of our approach, we implemented a working prototype and ran it against both a case study and a benchmark web application.
The experimental results confirm that our tool automatically discovers and executes a number of injection flaw attacks that are out of reach for state-of-the-art web scanners.

\end{abstract}

\section{Introduction}
\label{beagle:sec:intro}

Vulnerability testing aims at providing concrete evidence that a target software, i.e., the \emph{Application Under Testing}
(AUT), can execute some dangerous operations.
In general, the best possible evidence is a replicable test case that exploits the vulnerability.
Noticeably, vulnerabilities are often well understood and documented.
Several online databases, e.g., see NVD\footnote{https://nvd.nist.gov/}, CVE\footnote{https://cve.mitre.org/} and Exploit-DB\footnote{https://www.exploit-db.com}, contribute to a shared and up-to-date knowledge base.
Nonetheless, myriads of vulnerabilities persist for years after they have been documented.

Commonly a vulnerability is caused by very few lines of code (often only one).
Checking the presence of some dangerous instructions is only the first step.
Indeed, the internal logic of the application might prevent any execution that activates the vulnerability.
Thus, the ultimate evidence is an \emph{exploit}, i.e., an execution that triggers the vulnerability.
In most cases, finding an exploit is a challenging task requiring the expertise of a skilled penetration tester.
Such difficulty is mostly due to the necessity of understanding both the control and data flow of the AUT.
It becomes even more complex for event-based software where some side effects influence the execution flow.
GUI-based web applications are a common scenario. 

\begin{figure}[t]
\includegraphics[width=\columnwidth]{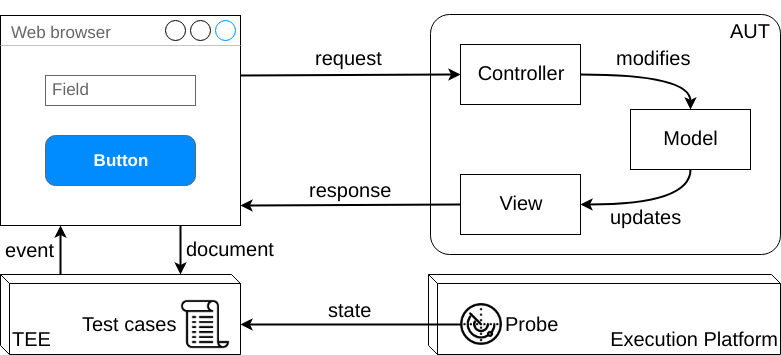}
\caption{The typical configuration for gray-box testing.}
\label{beagle:fig:guitesting}
\end{figure}

Very often a web application implements the MVC pattern.
The testing configuration for this case is shown in Figure~\ref{beagle:fig:guitesting}.
A \emph{Test Execution Engine} (TEE) stimulates
a web browser to interact with the AUT.
A web page is loaded and its components, e.g., buttons and text fields, triggered by some \emph{events}, e.g., clicking and typing.
A sequence of such events 
is a \emph{test} executed by the TEE.
A test causes one or more requests to be dispatched to the AUT controller module.
The controller interprets the requests and modifies the model, e.g., by assigning a value to a variable.
The model modification causes an update of the application view and, consequently, a response for the (browser of the) TEE.
Optionally, the TEE can be notified with some internal changes of the AUT state by a remote probe. 
When the observation of the internal state is not possible we call it \emph{black-box} testing.
Otherwise, we have a \emph{white-box} or a \emph{gray-box} TEE depending on the amount of information that the TEE can observe.

Unfortunately, the decoupling between the internal logic of the application and the interface events (such as in the MVC pattern) prevents a direct correlation between the vulnerability and the test that discloses it.
The reason is that the state of the AUT may be affected in a very different by each event.
For instance, typing text in a form may not affect the AUT until a button is pressed. 
On the other hand, inspecting the AUT code is also problematic.
Indeed, a web application may consist of several procedures that, according to the application state, are triggered by the requests of the browser.
As a consequence, human security experts commonly attempt to find a vulnerability test case by alternating interactions with the AUT and inspections of its code (e.g., see~\cite{OWASPtg}).

In principle, automated test case generation and execution frameworks can be very helpful: they can automatically scan the AUT employing a set of test cases (e.g., encoded as scripts)
and provide the human analyst with a list of vulnerabilities and the instructions for the exploit.
In practice, these tools are mainly used for a preliminary analysis phase, to collect information or test some common vulnerability patterns.
The main reason is that a TEE is as good as its test cases.
Indeed, if there is no guarantee that the tests cases reach the relevant corner cases, the risk of false negatives becomes critical.
Needless to say, the generation of useful test cases is extremely hard. We will show this later in the paper through a motivating example.

\smallskip
\noindent
In this paper, we present a novel technique for the automatic vulnerability test case generation and execution.
Our testing technique is based on a \emph{collaborative, co-evolutionary contract-driven} algorithm.
The fittest individuals in a population of tests are selected to generate an offspring for the next testing round.  
The \emph{fitness function} is based on a \emph{contract distance}, i.e.,~the minimum distance between the values generated by a test and the set of values that satisfy a \emph{contract}.
A contract is a sufficient condition for the test take an execution path in a procedure (associated with the contract).
The last contract to be satisfied is associated with the vulnerability itself.
Satisfying all the contracts of the procedures that the test executes eventually leads to triggering the vulnerability.

Since the calculation of the contract distance is a computationally hard problem, we to tackle it through a collaborative co-evolutionary algorithm  where one species, the \emph{test species}, evolves to find the target test, whereas $n$ additional species, the \emph{contract species}, evolve to approximate the call distances (one for each of the $n$ procedures implementing the AUT).

The result is a method that, starting from a generic specification of a vulnerability, automatically improves the test cases.
Although not guaranteed, the method often converges and reaches a vulnerability exploit.
We provide experimental evidence of this behavior (see below).

Although our technique has broader applicability, in this paper we explore its application on \emph{injection flaw} vulnerabilities, i.e., vulnerabilities where an attacker can maliciously interact with an interpreter through some crafted payload which is not correctly validated by the AUT.
In the last 15 years, this class of vulnerabilities has appeared continuously in the OWASP Top 10 vulnerabilities and, since 2010 it ranks first.\footnote{\url{https://www.owasp.org/index.php/Category:OWASP_Top_Ten_Project}}

We have developed \toolname{}, a prototype implementation of our technique. To assess the effectiveness of our technique we have run \toolname{} against two web applications. 
The first one is \casestudyapp{} (\emph{signup-confirm-welcome}), a minimal web application that implements an entry-preview-confirm pattern (akin those implemented by many websites to register new users) and suffers from a \emph{multi-step stored cross-site scripting} (XSS) vulnerability.
Detecting such a vulnerability is challenging for automated tools since it requires the knowledge of the logic of the AUT, i.e.,~the establishment of a ``session'' between the client and the server.
Moreover, our case study applies a filter that partially sanitizes the payload. Bypassing the filter is a further difficulty that often causes false negatives.
The second application is {WackoPicko}~\cite{Vigna10johnny}, a web application that has been specifically designed and used as a benchmark for penetration testing tools.
WackoPicko contains 8 injection flaws.
The results show that \toolname{} effectively detects all the vulnerabilities but one and, for each of them, it automatically generates an exploit test case.
Moreover, all exploit test cases are returned in a few minutes.
We also ran OWASP ZAP\footnote{\url{https://www.owasp.org/index.php/OWASP_Zed_Attack_Proxy_Project}},
w3af\footnote{\url{http://w3af.org/}}
and Vega\footnote{\url{https://subgraph.com/vega/index.en.html}},
three popular state-of-the-art web scanners,
against \casestudyapp{} and WackoPicko
They were unable to detect most of the vulnerabilities found by \toolname{}.

\smallskip
\noindent
\emph{Structure of the paper.} In Section~\ref{beagle:sec:motivation} we present a motivating example. 
Section~\ref{beagle:sec:overview} provides an overview of our approach, while Sections~\ref{beagle:sec:fitness} and~\ref{beagle:sec:genetic} detail our fitness function and co-evolutionary algorithm, respectively.
Then, in Section~\ref{beagle:sec:tool} we present our prototype \toolname{} and, in Section~\ref{beagle:sec:casestudy} we apply it to an experimental benchmark and we compare its performance against some state-of-the-art web scanners.
Finally, Section~\ref{beagle:sec:related} surveys on the related literature and Section~\ref{beagle:sec:conclusion} concludes the paper.


\section{Motivating example}
\label{beagle:sec:motivation}

In this section, we present \casestudyapp{}, a web application inspired to the \emph{multi-step stored XSS} scenario of~\cite{Vigna10johnny}.
\casestudyapp{} will serve us both as a motivating and working example. 
\casestudyapp{} is a web application that carries out a simple signup procedure.
Initially, the user is prompted with a single-field form where she has to insert a \emph{valid} username and then click a \emph{submit} button.
A valid username must have at least 6 characters, one of which being a digit.
Moreover, the web page applies a sanitization filter that removes the \verb!'! character from the input.
After submitting the form, the user is prompted with a confirmation page containing two links, i.e., \emph{confirm} and \emph{back}.
The first one points to a welcome page, while the second points back to the signup form.
The \emph{welcome} page displays the message \verb|Hello $user|, where \verb!$user! is
a variable containing the (sanitized) username inserted in the signup form. 

\begin{figure}[t]
\begin{center}
\includegraphics[width=\columnwidth]{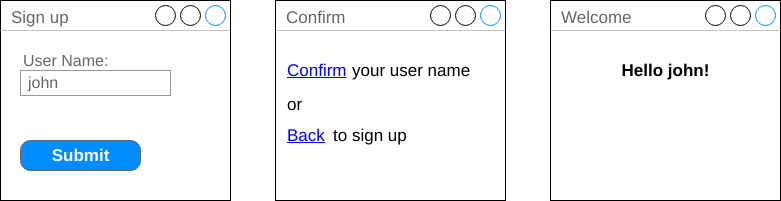}
\caption{The example web application}
\label{beagle:fig:webapp}
\end{center}
\end{figure}

A minimal PHP implementation of the web application described above is given in
Figure~\ref{beagle:fig:php}.
The implementation consists of three procedures, i.e., \verb!signup!, \verb!confirm!,
and \verb!welcome!.
\verb!signup! returns the HTML code of the signup form while \verb!welcome! returns the welcome message containing the username.
The code of \verb!confirm! requires more attention.
It consists of two, alternative branches that are executed according to the truth value assumed by a condition.
The condition is satisfied when the submitted username passes three checks,
i.e., $(i)$ it is not null, $(ii)$ it is at least 6 characters long and $(iii)$
it contains at least one digit.
If so, a filter that removes all \verb!'! characters is applied and the user mane is
stored in the session variable \verb!name!.
Eventually, the procedure returns the HTML code of the confirmation page.
Instead, when the condition is not satisfied, the \verb!confirm! procedure redirects the browser to \verb!signup!.

\begin{figure}[t]
\begin{lstlisting}[language=PHP,numbers=none,breaklines=true]
// signup.php
echo 'User Name:<br/>
      <form action="confirm.php" method="GET">
        <input type="text" name="payload">
        <input type="submit">
      </form>';

// welcome.php
session_start();
\$name = \$_SESSION['name'];
echo 'Hello \$name!';

// confirm.php
session_start();
if (isset(\$_GET['payload']) 
    && preg_match('/(\w){6,}/', \$_GET['payload'])
    && preg_match('/(.*\d.*)/', \$_GET['payload'])) {

 \$_SESSION['name'] = preg_replace('\'', '', \$_GET['payload']));
 echo '<a href="welcome.php">Confirm</a>
       <br/>or<br/>          
       <a href="signup.php">Back</a>';
} else header('Location: signup.php');
\end{lstlisting}
\caption{The PHP code for the web application.}
\label{beagle:fig:php}
\end{figure}

The signup application suffers from a stored XSS vulnerability.
A trivial way to exploit the vulnerability is to insert the username
\begin{center}
\verb!<script>alert(1)</script>!
\end{center}

\noindent
and then proceed to the welcome page.
In terms of GUI events, the sequence of actions to obtain the test is
\begin{enumerate}
\item click on the username field;
\item type \verb!<script>alert(1)</script>!;
\item click on button submit;
\item click on link confirm.
\end{enumerate} 

Sometimes, finding a vulnerability may be even simple for a human being.
Nevertheless, automatic vulnerability scanners usually fail to spot out a stored XSS vulnerability such as the previous one.
One of the main reasons is the size of the search space.
A measure of the search complexity can be given in terms of probability of success of a random tester.
Let us assume that the type and order of the events, i.e., click-type-click-click, is given.
Any randomly generated event $e_i$ (with $i \in {1, \ldots, 4}$) has a certain
probability of being the correct one.
The probability $Q_i$ is given by the number of correct events (e.g., the clicks that hit
a specific button) over the total number of possible events.
Thus, the overall probability $Q$ of a trace is the product of the probabilities of the events it consists of.

To simplify, let us assume that all pages are displayed on a $128 \times 128$ pixel's
area and that all the buttons, links and text fields are $64 \times 32$ pixel's
rectangles.
Also, we set the length of the text input to (exactly) $30$
ASCII\footnote{We only considered the $94$ printable characters.} characters.
In order to reveal the vulnerability, a username must be of the form
\texttt{<script>alert($a_1 a_2 a_3 a_4 a_5 a_6$)</script>} where each $a_i$ is either a
digit or the \textquotesingle{} symbol, and at least one of them is a digit.
Thus, the number of possible combinations is
$\sum_{i=1}^6 10^i \cdot \binom{6}{6-i} = 1771560$.
Under these settings we have
$
Q_1 \!=\! Q_3 \!=\! Q_4 \!=\! \frac{64 \cdot 32}{128^2} \!=\! \frac{1}{8}$
and $Q_2 \!=\! \frac{14640}{94^{28}} \!\cong\! 1.13 \times 10^{-53}$
which results in $Q = Q_1 \cdot Q_2 \cdot Q_3 \cdot Q_4 \cong 2.21 \times 10^{-56}$.

The previous example is extremely hard for a random tester.
Trivially, the main reason is $Q_2$, i.e., the text typing action.
This is not surprising as the number of possible strings grows exponentially with
the string length.

To overcome this issue, several tools rely on heuristics that reduce the search space.
Some of them, e.g., OWASP ZAP, inject predefined payloads into the web pages of the AUT. 
This method is effective to detect vulnerabilities that are exploited through unfiltered
payloads and that do not depend on the application state (e.g., Reflected XSS).
The same technique is not effective against custom filters or stateful vulnerabilities
(e.g., Stored XSS).
It is important to notice that the filter applied to the username in our example creates
a discrepancy between the input text and the actual payload.
This typically interferes with the test execution and makes it more difficult.
The reason is that simple heuristics can hardly generate a test with both $(i)$ input values that pass the custom checks and manipulations in the code of the AUT and $(ii)$ stimulates the execution flow that leads to the vulnerable instructions with a state configuration that exploits them.


\section{Overview}
\label{beagle:sec:overview}

Out TEE relies on an EA to automatically generate the test that reveals a target vulnerability.
The purpose of an EA is to (efficiently) converge to the minimum of a target function called \emph{fitness}.
The function $f$ we are interested in must assign a fitness value to any test $t$ that is executed on the AUT.
In particular, we want $f$ to be \emph{positive} i.e.,  $\forall t. f(t) \geq 0$ and such that $f(t) = 0$ if and only if $t$ is an exploit of the target vulnerability.

The fitness function is a cornerstone of every EA as their effectiveness and efficiency depend on it.
For instance, fitness functions having many local minima may be challenging to deal with, as the algorithm could get stuck around a suboptimal solution.
Als,o step functions are problematic: an EA might spend a considerable amount of time to explore constant plateaus, where its performance is similar to a random search.
This behavior is known as \emph{stagnation}.
Although the EAs can deal with a step function, the size of plateaus should be reduced as much as possible to avoid stagnation.

Even more problematic, defining a fitness function is often a difficult task.
This happens, for instance, when a problem has multiple dimensions or when the characterization of the search space is complex.
Our method needs a fitness function that, given a test $t$, returns a measure of how far $t$ is from the execution of the target vulnerability.
Each test stimulates the AUT to execute the sequence of its procedures.
Each procedure works on a set of parameters that drive its execution.
In Section~\ref{beagle:sec:vulnerability} we introduce a notion of \emph{contract}.
Briefly, a contract is a specification over the parameters of an invocation.
Intuitively, whenever a test satisfies the contract of a particular procedure, it gets closer to finding the vulnerability.
Otherwise, it cannot proceed further, and a better test must be generated.
In Section~\ref{beagle:sec:fitness} we define a distance function that estimates how far a test is from satisfying a contract.


We devote Section~\ref{beagle:sec:fitness} to the (incremental) definition of the fitness function.
One of the steps of its calculation consists of a computationally hard minimization problem in a multidimensional space. 
Since the fitness function is computed several times for each cycle of the EA, applying an analytical, exact method would lead to an extremely inefficient implementation.
We avoid this issue by using a \emph{collaborative co-evolutionary algorithm} (CCEA).
Briefly, some ancillary EAs co-evolve with the main one to provide a good approximation of the distance functions required for the calculation of the test fitness function. 

\begin{figure}[t]
\begin{lstlisting}[language=Pascal,mathescape]
Begin Species$_i$
  pop$_i$ := initialize();
  repeat
    fitness$_i$(pop$_1$,$\ldots$, pop$_n$);
    parents$_i$ := select$_i$(pop$_i$);
    offspring$_i$ := crossover$_i$(parents$_i$);
    pop$_i$ := mutate$_i$(offspring$_i$);
  until termination_condition();
End
\end{lstlisting}
\caption{A generic co-evolutionary algorithm.}
\label{beagle:fig:coevolutionary}
\end{figure}

The structure of an $n$-species CCEA is reported in Figure~\ref{beagle:fig:coevolutionary}.
The main difference w.r.t. a traditional EA is that $n$ species evolve concurrently.
Each species $i \in [1,n]$ has its population as well as fitness, selection, crossover and mutation functions.
However, the calculation of the fitness for the $i-$th species may also depend on the other populations. 
The design of our CCEA is presented in Section~\ref{beagle:sec:genetic}.

\section{Fitness function}
\label{beagle:sec:fitness}

In this section, we define the fitness function for our EA-based TEE.
We start by identifying the success conditions of a test.
Then we define a step function and we refine it through a correction factor.

\subsection{Vulnerability specification and success conditions}
\label{beagle:sec:vulnerability}

A vulnerability specification defines the conditions that trigger a vulnerability and, thus, identify a successful test.

\begin{definition}
\label{def:vulnspec}
A \emph{vulnerability specification} is a pair $V = \langle S(\bar{x}), C(\bar{x}) \rangle$ where $S$ is a sequence of instructions parametrized over the free variables $\bar{x} = x_1, \ldots x_n$ and $C$ is a predicate on $\bar{x}$. 
We call $S$ the \emph{vulnerability signature} and $C$ the \emph{(vulnerability) contract}.
\end{definition}

\begin{table}[t]
\caption{The contract specification syntax.}
\label{tab:vulnsyntax}
\begin{tabular}{l @{\,::=\:} l}
$C$ & $A_P \orsyn \neg C \orsyn C \vee C  \orsyn C \wedge C$ \\
$A_P$ & $B_A$ \orsyn $B_S$ \orsyn $B_B$ \\
$B_A$ & $A_E > A_E$ \orsyn $A_E < A_E$ \orsyn $A_E = A_E$\\
$A_E$ & $A_V$ \orsyn $A_E + A_E$ \orsyn $A_E - A_E$ \orsyn $A_E \cdot A_E$ \orsyn $A_E / A_E$ \orsyn \verb|len|$(Var)$  \\
$A_V$ & $Num$ \orsyn $Var$ \\
$B_S$ & $Var \in R_E$ \\
$R_E$ & $Str \!\orsyn Var \!\orsyn \!R_E . R_E\! \orsyn R_E\! +\! R_E \orsyn\! R_E^\ast$ \\
$B_B$ & $\mathtt{true}$ \orsyn $\mathtt{false}$ \orsyn $Var$
\end{tabular}
\end{table}

A contract $C$ is either an atomic predicate $A_P$ or composition of sub-predicates through the standard logic operators.
Atomic predicates are defined over arithmetic ($B_A$), string ($B_S$) or boolean ($B_B$) expression.
Comparisons between arithmetic expressions $A_E$ are standard.
Also, arithmetic expressions $A_E$ are standard.
Only note that an arithmetic expression can also be the length of a string variable $x$, i.e., \verb|len|$(x)$.
An atomic expression $A_V$ is either a number $n \in Num$ or a variable $x \in Var$.
String expressions ($B_S$) reduce to checking whether a string variable belongs to the language generated by a regular expression $R_E$.
Regular expressions are constant strings (delimited by \verb!"! and belonging to $Str$), variables or compositions of sub-expressions through the sequence $.$, choice $+$ or recursion (Kleene-star) operators.
Finally, boolean atoms ($B_B$) are either constants or variables.
We feel free to use parentheses to avoid ambiguity and we introduce some syntactic sugar, e.g. $\mathtt{[a{-}Z]}$ for \verb!"a"!$+$\verb!"b"!$+ \ldots +$\verb!"Z"!, $\Sigma$ for any character, $R^n$ for $R.R.\cdots.R$ (repeated $n$ times) and $x = s$ for $x \in s$ (being $s$ a constant string).

\begin{example}
\label{ex:specvuln}
Consider the scenario of Section~\ref{beagle:sec:motivation}.
We specify the XSS vulnerability as 
\begin{small}
\begin{center}
$\langle$\verb!echo! $x, x \in \Sigma^*.$\verb!"<script>alert("!$.R.$\verb!")</script>"!$.\Sigma^*\rangle$
\end{center}
\end{small}
\noindent
where $R = (0 + \mathtt{[1{-}9]}.\mathtt{[0{-}9]}^*) +$\textquotesingle$.(\mathtt{[0{-}9]}+\mathtt{[a{-}Z]})^*.$\textquotesingle, i.e., $R$ is either a natural number or a text (an alphanumeric sequence delimited by \textquotesingle).\footnote{Notice that the evaluation of the HTML tags is case insensitive, e.g., \verb!<sCRiPt>! can be used in place of \verb!<script>!. 
For brevity, we do not include this in the example specification.}
\end{example}

Below we introduce some definitions necessary to provide the characterization of a successful test.

\begin{definition}
\label{def:aut}
An \emph{AUT} is a finite set $A = \{p_1(\bar{x}_1), \ldots, p_n(\bar{x}_n)\}$ where each $p_i$ is a procedure and $\bar{x}_i$ is the list of the formal parameters of $p_i$.
\end{definition}

\begin{definition}
\label{def:trace}
Given an AUT $A = \{p_1(\bar{x}_1), \ldots, p_n(\bar{x}_n)\}$, the \emph{execution trace} of a test $t$, in symbols $\sigma(t) \in A^k$, is a finite sequence of procedures $p_{j_1}(\bar{v}_1), \ldots, p_{j_k}(\bar{v}_k)$ (with their actual parameters $\bar{v}_i$) invoked during the execution of $t$. 
\end{definition}

\begin{example}
\label{ex:trace}
Consider the AUT and the four-actions test $t$ of Section~\ref{beagle:sec:motivation} again.
The execution trace of $\sigma(t)$ is 
\[
\left.\begin{array}{l}\verb!signup.php()!,\\
\verb!confirm.php(<script>alert(1)</script>)!, \\
\verb!welcome.php(<script>alert(1)</script>)!
\end{array}\right.
\]
\vspace{-26pt}

\end{example}

\begin{definition}
Given a vulnerability specification $V = \langle S(\bar{x}), C(\bar{x}) \rangle$, we say that $p(\bar{y})$ is a \emph{target procedure} if the body of $p$ contains the sequence of instructions $S(\bar{e})$ (where $\bar{e}$ are arbitrary expressions).
\end{definition}

\begin{example}
According to the vulnerability specification of Example~\ref{ex:specvuln} \verb!welcome.php! (see Figure~\ref{beagle:fig:php}) is a target procedure since it contains the instruction \lstinline{echo 'Hello $name'}.
\end{example}

\begin{definition}
A test $t$ is \emph{successful} with respect to a vulnerability specification $\langle S(\bar{x}), P(\bar{x}) \rangle$  if and only if $(i)$ $\sigma(t) = p_1(\bar{v}_1), \ldots, p_k(\bar{v}_k)$, $(ii)$ $p_i(\bar{y})$ is a target procedure (for some $i \in [1, \ldots, k]$) and $(iii)$ the invocation $p_i(\bar{v}_i)$ executes $S(\bar{w})$ such that $P(\bar{w})$ holds.
\end{definition}

In words, a successful test causes the invocation of a sequence of procedures, one of which ($p_i$) being a target procedure.
Moreover, the execution of $p_i$ reaches the sequence of instructions $S$ under a context, i.e., the values $\bar{w}$, that complies with the vulnerability specification.

\begin{example}
The test $t$ of Example~\ref{ex:trace} is successful with respect to the vulnerability specification of Example~\ref{ex:specvuln}.
Also a test leading to the execution of the instruction \lstinline{echo "Hello <script>alert('xss')</script>!"} would be successful. 
However, notice that such a test cannot occur in the AUT of our working example.
The reason is that the filter in \texttt{confirm.php} removes the \textquotesingle{} symbols and $P($\lstinline{"Hello <script>alert(xss)</script>!"}$)$ does not hold (since \verb!xss! is neither a number nor a string).
\end{example}

Whenever a test is not successful we call it \emph{partial}.
As stated in Section~\ref{beagle:sec:overview}, a fitness function $f$ must ensure that $f(t) = 0$ if and only if $t$ is successful.
In Section~\ref{beagle:sec:stepfun} we define a distance function for partial tests.

\subsection{Call distance function for partial tests}
\label{beagle:sec:stepfun}

Given a vulnerability specification $V = \langle S(\bar{x}), P(\bar{x}) \rangle$, we define the distance of a test $t$ with respect to $V$, in symbols $d_V(t)$ (or simply $d(t)$ when clear from the context), to be the minimum distance between the procedures invoked by $t$ and the target procedures in the call graph of the AUT.

\begin{example}
Consider again the application of Section~\ref{beagle:sec:motivation} and the vulnerability specification of Example~\ref{ex:specvuln}.
Now assume that the only target procedure is \texttt{welcome.php}.
The call graph\footnote{Notice that here we are only limited to the AUT procedures, while, in general, the term ``call graph'' has a slightly different meaning.}, annotated with the distance of each node, is shown in Figure~\ref{beagle:fig:example-cg}.

\begin{figure}
\begin{center}
\includegraphics[width=\columnwidth]{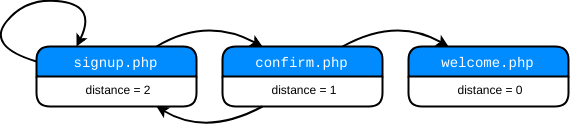}
\end{center}
\caption{The call graph of the example AUT.}
\label{beagle:fig:example-cg}
\end{figure}

The target \texttt{welcome.php} is at distance $0$ and it can be reached only through a link in \texttt{confirm.php} which, thus, has distance $1$.
Finally, \texttt{signup.php} has distance $2$ as it can reach \texttt{confirm.php}.

\noindent
Now consider the test $t$ that $(i)$ clicks on the user name field, $(ii)$ types \verb!john42!, $(iii)$ clicks on \emph{Submit}, and $(iv)$ clicks on \emph{Back}.
The test compiles the form with a valid user name and, through the submit button, moves from \texttt{signup.php} to \texttt{confirm.php}.
Then it clicks on \emph{back} and returns to \texttt{signup.php}.
In this case $d(t) = 1$. 
\end{example}

According to the above example $d(t) = 0$ does not imply that $t$ is successful.
As a matter of fact $d(t) = 0$ occurs whenever $\sigma(t)$ contains an invocation $p(\bar{v})$ where $p(\bar{x})$ is a target procedure. 
Nevertheless, $p(\bar{v})$ might not trigger the vulnerability, e.g., if the execution flow does not reach the vulnerable instructions.
Thus we recur to the following refinement.

\begin{definition}{(Call distance)}
\label{def:calldistance}
Let $t$ be a test such that $\sigma(t) = p_1(\bar{v}_1), \ldots, p_n(\bar{v}_n)$ and let $d(p_i)$ the distance function given above, then we define the \emph{call distance} $\delta$ as
\[
\delta(t) = \min_{p_i \in \sigma(t)} d(p_i) + \left\{\begin{array}{l}
0 \textnormal{ if } t \textnormal{ is successful} \\
1 \textnormal{ otherwise}
\end{array}\right. 
\]
\vspace{-32pt}

\end{definition} 

The $\delta$ function is a candidate for the fitness function.
Indeed, it is positive, i.e., $\forall t.\delta(t) \geq 0$ and $\delta(t) = 0$ if and only if $t$ is successful, and monotonic, i.e., $\forall t,t'. \sigma(t) \subseteq \sigma(t') \rightarrow \delta(t) \leq \delta(t')$.
However, $\delta$ is a step function mapping each test to a plateau.
The plateau corresponds to all the possible values assumed by the parameters of the procedure $p_i$ in Definition~\ref{def:calldistance}.
As discussed above, step functions reduce the EA to a random search.
We propose the following example to clarify this aspect.

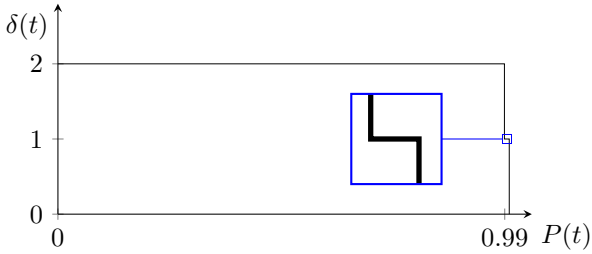
\begin{figure}[t]
\begin{tikzpicture}[spy using outlines={rectangle,lens={scale=3}, size=0.4cm, connect spies}]
\begin{axis}[%
    ,xlabel=$P(t)$
    ,ylabel=$\delta(t)$
    ,y=1cm
    ,x=6cm
    ,axis x line = bottom,axis y line = left
    ,xmax=1.05
    ,xtick={0,0.99}
    ,ytick={0,1,2}
    ,ymax=2.8 
    ]
\addplot[const plot, no marks, very thin] coordinates {(0,2) (0.9893,1) (1,0)};
\end{axis}
\spy[blue,size=1.2cm,magnification=10] on (5.97,1) in node at (4.5, 1);
\end{tikzpicture}
\caption{Step function induced by $\delta(t)$.}
\label{beagle:fig:stepfun}
\end{figure}

\begin{example}
\label{ex:stepfun}
We assume that all the tests have the structure described in Section~\ref{beagle:sec:motivation}.
Given a random test $t$, we compute the probability $P$ of $\delta(t) = 2$, $\delta(t) = 1$ and $\delta(t) = 0$.

\noindent
$\delta(t) = 2$ happens when $t$ never leaves \texttt{signup.php}.
The probability of such event is
$$P(\delta(t) = 2) = 1 - P(\textnormal{click}_{\textnormal{user}}) \cdot P(\textnormal{type}_{\textnormal{valid}}) \cdot P(\textnormal{click}_{\textnormal{submit}})$$
\noindent
$P(\textnormal{type}_{\textnormal{valid}})$ is non trivial.
Assuming a maximum string length of 30 characters we have
\[
P(\textnormal{type}_{\textnormal{valid}}) = \sum\limits_{n=6}^{30} \frac{1 - \left(\frac{84}{94}\right)^n}{30} \cong 0.68
\]
As a matter of fact, $P(\textnormal{type}_{\textnormal{valid}})$ amounts to the sum of the probabilities of generating a valid input of a given length.
We assume that each length between 1 and 30 has the same probability, i.e., $1/30$.
The probability of string of length $n \geq 6$ is the dual of that of an invalid one, i.e., a string only consisting of the $84$ non-digit characters.
Summing up, we have $P(\delta(t) = 2) \cong 0.9893$.

\noindent
$\delta(t) = 0$ happens when $t$ is successful.
Thus, the probability $P(\delta(t) = 0)$ is equal to
$$P(\textnormal{click}_{\textnormal{user}}) \cdot P(\textnormal{type}_{\textnormal{valid} \wedge \textnormal{xss}}) \cdot P(\textnormal{click}_{\textnormal{submit}}) \cdot P(\textnormal{click}_{\textnormal{confirm}})$$
Assuming that $P(\textnormal{type}_{\textnormal{valid} \wedge \textnormal{xss}})$ is the probability of generating a random 30-characters string of Section~\ref{beagle:sec:casestudy}, we obtain 
$$P(\delta(t) = 0) \cong 2.207 \times 10^{-56}$$

\noindent
Finally, $P(\delta(t) = 1) = 1 - P(\delta(t) \neq 1) \cong 0.0107$. Figure~\ref{beagle:fig:stepfun} depicts the step function induced by the probability distribution of $\delta(t)$.
The enlarged area shows the step for $P(\delta(t) = 1)$.
\end{example}

Example~\ref{ex:stepfun} highlights that the call distance function $\delta(t)$ is very close to a constant function.
Hence, to obtain a reasonable fitness function we need to correct the it.

\subsection{Correction to the call distance}
\label{beagle:sec:correction}

To refine $\delta(t)$ we introduce a correction factor.
Intuitively, the correction factor is the distance of a test $t$ such that $\delta(t) = n$ (with $n > 0$) from the \emph{closest} test $t'$ such that $\delta(t') = n-1$.

\smallskip
\noindent
\textbf{Call contracts.}
We start from the call graph and the call distances of the AUT as discussed in Section~\ref{beagle:sec:stepfun}.
For each procedure $p$ such that $d(p) = n > 0$ there must be at least one $p'$ such that $(i)$ $d(p') = n-1$ and $(ii)$ $p$ contains an invocation to $p'$.
Since the control flow of $p$ is driven by the data, e.g., inputs and session variables, we aim at identifying (and put in a contract) the conditions under which $p$ invokes $p'$.

By using the same contract specification language of Table~\ref{tab:vulnsyntax} we define a \emph{call contract} $C_p$ for each procedure $p$.
In general, a call contract represents a sufficient condition for $p$ to invoke a $p'$ such that $d(p) = d(p') + 1$.

\begin{example}
\label{ex:contracts}
Consider again the procedure \verb!confirm.php! in Table~\ref{beagle:fig:php}.
It can behave in two different ways depending on the evaluation of the \verb!if! guard: either it shows the confirmation message or it redirects to \verb!signup.php!.
The confirmation message contains a link to \verb!welcome.php!.
Since $d(\verb!welcome.php!) = 0$ and $d(\verb!signup.php!) = 2$ the call contract must imply that the \verb!if! guard evaluates to true.
A suitable call contract $C_{\verb!confirm.php!}$ is
\[
\begin{array}{c}
\verb!$_GET[`payload']! \in \Sigma^*.[0{-}9].\Sigma^* \\
\land $\quad$ len(\verb!$_GET[`payload']!) \geq 6
\end{array}
\]

Moreover, since \verb!signup.php! has no parameters, we have that $C_{\verb!signup.php!} = \mathtt{true}$.
\end{example} 

We also attach a call contract to each target procedure $p$, i.e., $d(p) = 0$.
The only difference is that the call contract of a target procedure is a sufficient condition to trigger the vulnerability.  

\begin{example}
Consider again the procedure \verb!welcome.php! in Figure~\ref{beagle:fig:php}.
A call contract ensuring the activation of the XSS vulnerability is
\begin{small}
\[
\verb!$_GET[`payload']! \in \verb!"<script>alert("!.R.\verb!")</script>"!
\]
\end{small}
\noindent
where $R$ is defined as in Example~\ref{ex:specvuln}.
\end{example} 

\smallskip
\noindent
\textbf{Contract distance.}

Intuitively, the reason behind the step structure of $\delta$ is that it only considers which procedures are invoked and it neglects what data they process.
In this section, we introduce a data distance function based on the call contracts.

\begin{definition}
Let $\bar{v} = (v_1 : \mathbb{T}_1, \ldots, v_n : \mathbb{T}_n)$ and $\bar{w} = (w_1 : \mathbb{T}_1, \ldots, w_n : \mathbb{T}_n)$ be two vectors of typed values.
Let us assume that the distance functions $| \cdot, \cdot |_i : \mathbb{T}_i \times \mathbb{T}_i \rightarrow \mathbb{N}$ are defined.
We define the distance between $\bar{v}$ and $\bar{w}$ as
\[
\Vert \bar{v}, \bar{w} \Vert = \sum_i |v_i, w_i|_i
\]
That is, $\Vert \cdot,\cdot \Vert$ is a \emph{Manhattan distance}.\end{definition}

\smallskip
Since our contract language is defined on three types, i.e., booleans ($\mathbb{B}$), integers ($\mathbb{Z}$) and strings ($\mathbb{S}$), here we consider three distance functions, i.e., $|\cdot|_\mathbb{B}$, $|\cdot|_\mathbb{Z}$ and $|\cdot|_\mathbb{S}$.
The first two are trivially defined as
\[
|a,b|_\mathbb{B} = \left\{\begin{array}{l}
0 \textnormal{ if } a = b\\
1 \textnormal{ otherwise}
\end{array}\right.
\qquad
|n,m|_\mathbb{Z} = |n - m|
\]

\noindent
while for $|\cdot|_{\mathbb{S}}$ we use the \emph{Levenshtein distance}~\cite{levdist}.


\begin{definition}
Given a call contract $C$, the \emph{contract distance} function $\gamma_c$ of a vector $\bar{v}$ is defined as
\[
\gamma_c(\bar{v}) = \min_{\bar{w} : C(\bar{w})} \Vert \bar{v},\bar{w}\Vert 
\]
In words, $\gamma_c(\bar{v})$ is the smaller distance between $\bar{v}$ and one of the vectors that satisfy $C$. 
\end{definition}

\begin{example}
\label{ex:contractbase}
Consider the following contract.
\[
C = \begin{array}{c}
\verb!$_GET[`payload']! \in \Sigma^*.[0{-}9].\Sigma^* \\
\land $\quad$ len(\verb!$_GET[`payload']!) \geq \verb!$_GET[`y']!
\end{array}
\]
$C$ is a slightly modified version of the contract $C_{\verb!confirm.php!}$ of Example~\ref{ex:contracts}.
The only difference is that the length of \verb!payload! must be greater or equal to a parameter \verb!y! (rather then $6$).
Also consider the vector $\bar{v} = (\verb!payload! = \verb!"john"!, \verb!y! = 7)$.
Since $C(\bar{v})$ does not hold, $\gamma_c(\bar{v}) > 0$.
The vector $\bar{u} = (\verb!"john42"!, 6)$ satisfies $C$ and $\Vert \bar{v}, \bar{u} \Vert = |\verb!"john"!, \verb!"john42"!|_\mathbb{S} + |7 - 6| = 2 + 1 = 3$.
To state that $\gamma_c(\bar{v}) = 3$ we need to show that there exist no $w$ such that $C(w)$ holds and $\Vert \bar{v}, \bar{w} \Vert < 3$.
This follows from the observation that $7 - len(\verb!john!) = 3$. 
Notice that other vectors are at distance $3$ from $\bar{v}$, e.g., $\bar{w} = (\verb!"0john"!, 5)$.
\end{example}

In general, calculating a contract distance is computationally complex.
As a matter of fact, it can be reduced to an integer programming problem which is known to be NP-hard~\cite{Papadimitriou81integer}.
Thus, an analytical approach is not compatible with our assumptions.
In Section~\ref{beagle:sec:genetic} we show how we carry out this computation as part of our EA.

\smallskip
\noindent
\textbf{Fitness function.}

By using the contract distance defined above, we define the fitness function as follows.

\begin{definition}
\label{def:fitness}
The \emph{fitness function} of a test $t$ is defined as
\[
\varphi(t) = \delta(t) - \frac{1}{\gamma_c(\bar{v}) + 1}
\]
\noindent
where $C$ is the contract of the procedure $p$ such that $\delta(t) = d(p) [+ 1]$ (see Definition~\ref{def:calldistance}).
\end{definition}

The term $\frac{1}{\gamma_c(\bar{v}) + 1}$ is called the \emph{correction factor} of the fitness.
Since $\gamma_c(\bar{v}) \in [0, +\infty[$ the correction factor belongs to $]0, 1]$.

\begin{example}
Consider two tests $t$ and $t'$ that reach the same page $p$ such that $\delta(t) = \delta(t') = d(p) = 2$.
Now imagine that the call contract of $p$ is $C$ and that $t$ and $t'$ invoke $p(\bar{v})$ and $p(\bar{w})$ (respectively), where $C$, and $\bar{v}$ are as in Example~\ref{ex:contractbase}, while $\bar{w} = (\verb!c4rl!, 5)$.
For $t$ and $t'$, the fitness function assumes the following values.
\[
\varphi(t) = 2 - \frac{1}{3 + 1} = \frac{7}{4} \quad > \quad \varphi(t') = 2 - \frac{1}{1 + 1} = \frac{3}{2}
\]
In words, $t'$ is a better test than $t$.
\end{example}


\section{CCEA Design}
\label{beagle:sec:genetic}

Our CCEA has $k + 1$ species $S_0, \ldots, S_k$ where $k$ is the number of AUT procedures and, thus, that of the call contracts.
The \emph{test species} $S_0$ is for test case generation, whereas the \emph{contract species} $S_1, \ldots, S_k$ aim at efficiently approximating the contract distance functions (see Section~\ref{beagle:sec:correction}). 

\subsection{Test species}
\label{beagle:sec:testspecies}

The test species consists of individuals, i.e., chromosomes, that encode a test of the AUT (see below).
In general, we assume a test to consists of a sequence of events.
However, and without loss of generality, in the following we often refer to GUI events such as clicks and text inputs.

\noindent
\textbf{Encoding and initialization.}
We assume $E = \{e_1(\bar{x}_i), \ldots, e_m(\bar{x}_m)\}$ 
to be the set of the \emph{events} supported by the AUT.
Each event $e(\bar{x})$ consists of an event label $e$ and a list of typed parameters $\bar{x}$, e.g., click$(x:\mathbb{N}, y:\mathbb{N})$.
An \emph{action} is an instance of an event where the parameters are replaced with actual values, e.g., click$(10, 3)$.
A \emph{test chromosome} is a permutation of $k_1 + \ldots + k_m$ actions where $k_i$ are inputs provided to the CCEA.
Each $k_i$ indicates the number of actions referring to the event label $e_i$ that occur in the test chromosomes.
The test chromosomes are initialized by generating a random permutation of random actions.

\begin{example}
\label{ex:chromo}
In our working example, we have $E = \{\textnormal{click}(x:\mathbb{N},y:\mathbb{N}), \textnormal{type}(s:\mathbb{S})\}$.
Then, we define the test chromosomes $t_1$ and $t_2$ as

\noindent
\begin{tabular}{l @{$\,=\,$} l}
$t_1$ & click$(17,5)$, click$(51,42)$, type(\verb!"john"!), click$(6,6)$ \\
$t_2$ & click$(4,15)$, type(\verb!"c4rl"!), click$(1,22)$, click$(9,55)$
\end{tabular}

\noindent
Both $t_1$ and $t_2$ have $k_\textnormal{click} = 3$ and $k_\textnormal{type} = 1$.
\end{example}

\noindent
\textbf{Crossover and mutation.}
The crossover function operates on $\ell$ positions of the test chromosomes.
The $\ell$ actions are randomly chosen among the $k_1 + \ldots + k_m$ forming each chromosome.
Then, we apply a local crossover to the pair of actions having the same index in the two target chromosomes.
Notice that, since the elements of each chromosome are indexed according to their event label, we always perform a crossover between two actions having the same structure.
For each pair of actions, we apply a single-point crossover to their parameters.  

\begin{example}
Consider again $t_1$ and $t_2$ from Example~\ref{ex:chromo}.
Now assume that $\ell = 2$ and that the randomly chosen actions are the first (only) \emph{type} and the third \emph{click}.

\begin{center}
\begin{tikzpicture}[scale=0.5]
\edef\sizetape{0.7cm}
\tikzstyle{tmtape}=[draw,minimum size=\sizetape,minimum width={62pt}]

\begin{scope}[start chain=1 going right,node distance=-0.15mm]
    \node [on chain=1,tmtape] {\small click$(17,5)$};
    \node [on chain=1,tmtape] {\small click$(51,42)$};
    \node [on chain=1,tmtape] (A) {\small type(\verb!"john"!)};
    \node [on chain=1,tmtape] (B) {\small click$(6,6)$};
\end{scope}

\begin{scope}[shift={(0cm,-3cm)}, start chain=2 going right,node distance=-0.15mm]
    \node [on chain=2,tmtape] {\small click$(4,15)$};
    \node [on chain=2,tmtape] (AA) {\small type(\verb!"c4rl"!)};
    \node [on chain=2,tmtape] {\small click$(1,22)$};
    \node [on chain=2,tmtape] (BB) {\small click$(9,55)$};
\end{scope}

\draw[{Latex[length=3mm]}-{Latex[length=3mm]}, line width=1pt] (A) -- (AA);
\draw[{Latex[length=3mm]}-{Latex[length=3mm]}, line width=1pt] (B) -- (BB);
\end{tikzpicture}
\end{center}

Now imagine that the crossover points are $1$ and $0$, respectively.
This means splitting the \verb!"john"! and \verb!"c4rl"! after the first character and keeping the click coordinates (thus swapping them entirely). 
The resulting offspring is the following.

\begin{center}
\begin{tikzpicture}
\edef\sizetape{0.7cm}
\tikzstyle{tmtape}=[draw,minimum size=\sizetape,minimum width={62pt}]

\begin{scope}[start chain=1 going right,node distance=-0.15mm]
    \node [on chain=1,tmtape] {\small click$(17,5)$};
    \node [on chain=1,tmtape] {\small click$(51,42)$};
    \node [on chain=1,tmtape] (A) {\small type(\verb!"jorl"!)};
    \node [on chain=1,tmtape] (B) {\small click$(9,55)$};
\end{scope}

\begin{scope}[shift={(0cm,-1cm)}, start chain=2 going right,node distance=-0.15mm]
    \node [on chain=2,tmtape] {\small click$(4,15)$};
    \node [on chain=2,tmtape] (AA) {\small type(\verb!"c4hn"!)};
    \node [on chain=2,tmtape] {\small click$(1,22)$};
    \node [on chain=2,tmtape] (BB) {\small click$(6,6)$};
\end{scope}
\end{tikzpicture}
\end{center}
\vspace{-8pt}

\end{example}

An action of a test chromosome of the offspring mutates with a certain probability.
When a mutation occurs we randomly apply one of two possible modifications.
The first mutation randomly modifies the parameters of an action, e.g., by modifying a character in a string.
The second mutation swaps the position of the target action with another one in the test chromosome. 

\noindent
\textbf{Fitness and selection.}
The fitness of a test chromosome is computed by means of the function $\varphi$ (see
Definition \ref{def:fitness}).
Then we apply the \emph{(deterministic) tournament selection} where the winning condition is the lower value of $\varphi$.
The tournament selection works as follows.
First $k$ chromosomes are randomly chosen among the $n$ forming the current population.
Then their fitness is computed and the best chromosome wins the tournament, i.e., is selected.
Crossover and mutation (see above) are then applied to the winners of the tournaments to obtain the next generation offspring.

\subsection{Contract species}
\label{beagle:sec:contractspecies}

For each procedure $p_i$, associated with a contract $C_i$, we have a species $S_i$.
The population of $S_i$ consists of chromosomes that (encode vectors that) satisfy $C_i$.

\noindent
\textbf{Encoding and initialization.}
A \emph{contract chromosome} is represented by a vector $\bar{v}$.
An SMT solver~\cite{smt} generates the initial population.
Intuitively, we proceed as follows.
We start by encoding a contract $C$ into a corresponding SMT specification.
Then, we run the solver to check the satisfiability of the specification and, in case, to obtain a \emph{model}, i.e., a vector $\bar{v}$ of values that satisfy the specification and, thus, the contract $C$.
To generate a further chromosome, we repeat the satisfiability check after invalidating the model $\bar{v}$.
Invalidating $\bar{v}$ requires to add few extra clauses to the specification of $C$ (see Example~\ref{ex:sat} below).
We iterate this process until we obtain enough chromosomes for the initial population or the specification becomes unsatisfiable.
In the second case, we complete the population with copies of the existing chromosomes.
 
The SMT encoding is straightforward since all of the operators in our contract language are directly mapped into a corresponding SMT statement.
For the sake of presentation, here we only provide the intuition through an example.
We refer the interested reader to~\cite{z3str} for further details on the theory of strings and SMT.

\begin{example}
\label{ex:smt}
The following SMT specification corresponds to the contract $C(\verb!payload!,\verb!y!)$ of Example~\ref{ex:contractbase}.

\begin{minipage}{\columnwidth}
\begin{lstlisting}[language=LISP,numbers=left, morekeywords={declare-const, String, Int, str,in,re,++,*,len,range}]
(declare-const payload String)
(declare-const y Int)
(assert 
  (str.in.re payload
    (re.++ (re.* (re.range " " "~"))
      (re.++ (re.range "0" "9")
        (re.* (re.range " " "~"))))))
(assert (>= (str.len payload) y))
\end{lstlisting}
\end{minipage}

In words, the specification declares two constants, i.e., \texttt{payload} and \texttt{y}, of type string and integer (lines 1 and 2).
The constants encode the two parameters of the contract.
The specification consists of two assertions (line 3 and 8, respectively).
They encode the two sides of the conjunction $C$ consists of.
Briefly, the mapping between the terms appearing in $C$ and those of the SMT specification is the following.

\noindent
\begin{tabular}{r @{$\:\equiv\:$} l @{\hspace{30pt}} r @{$\:\equiv\:$} l}
\hline
\verb!str.in.re! & $\in$ & \verb!re.range " " "~"! & $\Sigma$ \\
\verb!re.++! & $.$ & \verb!re.range "0" "9"! & $[0-9]$ \\
\verb!re.*! & $^*$ & \verb!str.len! & $len$ \\
\hline
\end{tabular}

\end{example}

The SMT solver checks the satisfiability of the specification for a contract $C$.
In case of failure (\emph{unsat}) the contract cannot be respected by any invocation to the procedure.
Otherwise (\emph{sat}), the solver returns a model, i.e., a vector of values $\bar{v}$ such that $C(\bar{v})$ is satisfied.
Then we add the model vector to the initial population of the contract species.
To generate further vectors we re-submit the specification after invalidating the model generated so far.
To invalidate a model we add a SMT statement that negates it.

\begin{example}
\label{ex:sat}
The specification of Example~\ref{ex:smt} admits the solution \verb!payload = "7"!, \verb!y = !$0$.
If such model is returned by the solver, we add the statement

\begin{lstlisting}[language=LISP, morekeywords={not}]
(assert (not (= payload "7")))
(assert (not (= y 0)))
\end{lstlisting}
\vspace{-8pt}

to the specification of Example~\ref{ex:smt}.
The new specification is satisfied, e.g., by \verb!payload = "G?_9"!, \verb!y = !$2$. 
\end{example}

\noindent
\textbf{Crossover and mutation.}
We apply a single point crossover to the contract chromosomes.
Recall that each chromosome encodes a $k$-values vector and that all the chromosomes consist of values of the same type.
We pick a random position, and we swap the two parts of the chromosomes.
The crossed chromosomes form the offspring.

\begin{example}
\label{ex:cross}
Consider the two vectors $(\verb!"7"!,0)$ and $(\verb!"G?_9"!,2)$ corresponding to the model generated in Example~\ref{ex:sat}.
Assume that the crossover pointcuts them in the middle.
The result of the crossover is $\bar{w}_1 = (\verb!"7"!,2)$ and $\bar{w}_2 = (\verb!"G?_9"!,0)$.
\end{example}

A mutation amounts to applying (with a constant probability) a random modification to a random element of a vector in the offspring.
The mutation of integers and booleans is straightforward.
For the strings we randomly apply one of these three operations: deletion, insertion and modification.
Briefly, deletion removes the character in a random position of the string, insertion adds a random character in a random position of the string and modification replaces the character in a random position of the string (with a random one).

\begin{example}
Consider again the vector $\bar{w}_2$ of Example~\ref{ex:cross}.
The three vectors $(\verb!"G?9"!,0)$, $(\verb!"GA?_9"!,0)$ and $(\verb!"Gr_9"!,0)$ are possible mutations (over the first element) resulting from the application of the deletion, insertion and modification operations, respectively.
\end{example}
 
\noindent
\textbf{Fitness and selection.}
The fitness of a contract chromosome (encoding the vector) $\bar{w}$ belonging to the population of $S_i$ is computed through the formula

\begin{equation}
\label{eq:fitnessi}
\textnormal{fitness}_i(\bar{w}) = \left
\{\begin{array}{l l}
\Vert \bar{v}, \bar{w} \Vert & \textnormal{ if } C_i(\bar{w}) \textnormal{ holds} \\
+ \infty & \textnormal{ otherwise}
\end{array}\right.
\end{equation}

\noindent
where $\bar{v}$ is the vector that contains the values generated by the execution of a test $t$ (see Section~\ref{beagle:sec:testspecies}).
In particular, $t$ causes the invocation of $p_i(\bar{v})$ where $p_i(\bar{x})$ is the procedure associated with the contract $C_i$ and, thus, with $S_i$.
Again, the selection process is based on the tournament method where the lower fitness value is the tournament winning condition.
Notice that the $+\infty$ fitness in (\ref{eq:fitnessi}) ensures that vectors that violate $C_i$ always lose against the vectors that satisfy $C_i$. 

\begin{example}
We compute the fitness of the two vectors $\bar{w}_1 = (\verb!"7"!,2)$ and $\bar{w}_2 = (\verb!"G?_9"!,0)$ (from Example~\ref{ex:cross}) w.r.t. the contract $C$ and vector $\bar{v}$ (both from Example~\ref{ex:contractbase}).
Since $\bar{w}_1$ does not comply with $C$, fitness$(\bar{w}_1) = +\infty$.
Instead, fitness$(\bar{w}_2) = \Vert \bar{v}, \bar{w}_2 \Vert = |\verb!"john"!,\verb!"G?_9"!|_\mathbb{S} + |7,0|_\mathbb{Z} = 4 + 7 = 11$.
\end{example}

\section{\toolname{}}
\label{beagle:sec:tool}

In this section, we present our prototype implementation \toolname{}.
We first provide an architectural description together with an overview of the CCEA parameters gauging. Then we briefly discuss the current limitations and future developments.
Beagle is publicly available on GitHub\footnote{\url{https://github.com/beagle-team/beagle}}.

\subsection{Architecture}

\begin{figure}[t]
\includegraphics[width=\columnwidth]{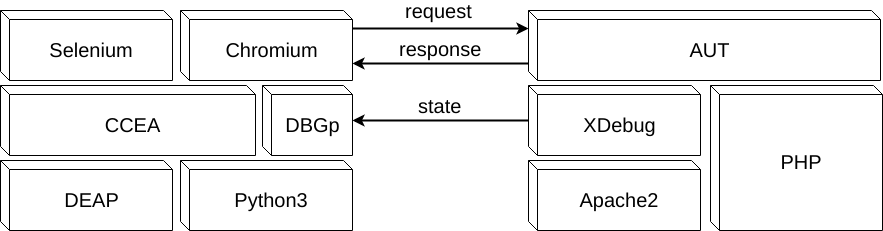}
\caption{Abstract architecture of \toolname{}.}
\label{beagle:fig:architecture}
\end{figure}

The abstract architecture of \toolname{} is depicted in Figure~\ref{beagle:fig:architecture}.
\toolname{} consists of two modules: the client-side TEE and the server-side AUT execution platform.
The TEE executes the CCEA presented in Section~\ref{beagle:sec:genetic} and it is based on DEAP~\cite{DEAP2012}, a framework for the implementation of generic EAs that we slightly extended to also support CCEA.
The client-side TEE relies on the Selenium\footnote{\url{https://www.seleniumhq.org/}} web driver and the (headless) Chrome web browser.\footnote{\url{https://developers.google.com/web/updates/2017/04/headless-chrome}}
Finally, the server-side AUT is a standard Apache web server used to monitor the AUT code execution through the XDebug\footnote{\url{https://xdebug.org/}} module.
XDebug monitors the call trace (i.e., procedures and parametersO of the AUT and transmits it to the TEE through the DBGp protocol.\footnote{\url{https://xdebug.org/docs-dbgp.php}}

\subsection{CCEA parameters gauging}

For an effective usage of EAs, a proper setting of the parameters is key.
The main parameters are the mutation and crossover probabilities, which we set to 0.95 and 0.06 respectively.

The mutation function depends on the target action.
The most complex mutation in our approach is the \emph{type} action mutation.
This action required careful gauging in order to avoid local minima in the EA.
When a type mutation happens, three different events can occur: $(i)$ it removes a random character from the string, $(ii)$ it inserts a randomly generated character, or $(iii)$ it modifies a random character into a new one.

Another important parameter to set is the maximum number of generations.
A \toolname{} test stops it reaches fitness 0,
however, as any stochastic technique, it might fail at reaching its target in a reasonable time.
It is important to set an upper bound to the computation that stops the algorithm when it stagnates for too long and proceeds to create a new test case that might succeed.

\subsection{Limitations and improvements}

Since \toolname{} is a prototype, it inevitably suffers from some limitations.
The main one is the lack of support for the generation of call contracts which must therefore manually written.
Although not complex and done on a routine basis (e.g., in contract-driven development~\cite{meyercdd}), manual contract writing is tedious and error-prone.
In principle call contracts can be automatically inferred through, e.g., a \emph{weakest preconditions} calculus~\cite{wpcal}. Unfortunately, no suitable implementation of a weakest preconditions calculus for PHP seems to be publicly available.

Another aspect to be discussed is the event/control system representation.
In \toolname{}, we abstract from the specific controls of the AUT by considering generic GUI stimuli.
For instance, our click events are characterized by the screen coordinates.
Other methods may provide concrete advantages.
For instance, a DOM-based approach can identify the controls that appear on a web page and directly stimulate them.
However, our representation can also support embedded elements, e.g., flash components, which are not available if the DOM abstraction is used.


\section{Experimental evaluation}
\label{beagle:sec:casestudy}

We executed \toolname{} on a dual-core, 2.6GHz vCPU, 4GB RAM virtual host provided by DigitalOcean.\footnote{\url{https://blog.digitalocean.com/introducing-high-cpu-droplets/}}
In such environment the average speed of \toolname{} is $9.44$ $g/s$ (generations per second).

An execution of \toolname{} consists of 10 worker threads running in parallel.
Each worker tests the AUT independently from the others.
A worker terminates when either it generates a successful test or it reaches the 50000th generation.
In the second case the worker results is negative, i.e., the best test generated by the worker is partial.
\toolname{} terminates with a negative result when all its workers do so.

\subsection{Application to the motivating example}

\begin{figure}[t]
\begin{tikzpicture}
\begin{axis}[%
    ,xlabel=$n\,(\times\!10^4)$
    ,ylabel=$\varphi(t^*_n)$
    ,x label style={at={(axis description cs:0.5,-0.1)},anchor=north}
    ,y label style={at={(axis description cs:-0.1,.5)},rotate=90,anchor=south}
    ,xmax=5
    ,y=1cm
    ,ymin=0
    ,xmin=0
    ,xtick={1,2,3,4,5}
    ,ytick={0,1,2,3}
    ,ymax=3.8 
    ]
\addplot+[mark=none, const plot, ultra thick] table [x expr=\thisrow{X}/10000, y expr=\thisrow{Y}] {./tests/case-study/digest_test3.csv};
\addplot+[mark=none, const plot] table [x expr=\thisrow{X}/10000, y expr=\thisrow{Y}] {./tests/case-study/digest_test0.csv};
\addplot+[mark=none, const plot] table [x expr=\thisrow{X}/10000, y expr=\thisrow{Y}] {./tests/case-study/digest_test1.csv};
\addplot+[mark=none, const plot] table [x expr=\thisrow{X}/10000, y expr=\thisrow{Y}] {./tests/case-study/digest_test2.csv};
\addplot+[mark=none, const plot] table [x expr=\thisrow{X}/10000, y expr=\thisrow{Y}] {./tests/case-study/digest_test4.csv};
\addplot+[mark=none, const plot] table [x expr=\thisrow{X}/10000, y expr=\thisrow{Y}] {./tests/case-study/digest_test5.csv};
\addplot+[mark=none, const plot] table [x expr=\thisrow{X}/10000, y expr=\thisrow{Y}] {./tests/case-study/digest_test6.csv};
\addplot+[mark=none, const plot] table [x expr=\thisrow{X}/10000, y expr=\thisrow{Y}] {./tests/case-study/digest_test7.csv};
\addplot+[mark=none, const plot] table [x expr=\thisrow{X}/10000, y expr=\thisrow{Y}] {./tests/case-study/digest_test8.csv};
\addplot+[mark=none, const plot] table [x expr=\thisrow{X}/10000, y expr=\thisrow{Y}] {./tests/case-study/digest_test9.csv};
\end{axis}
\end{tikzpicture}
\caption{Application to the case study.}
\label{beagle:fig:expcasestudy}
\end{figure}
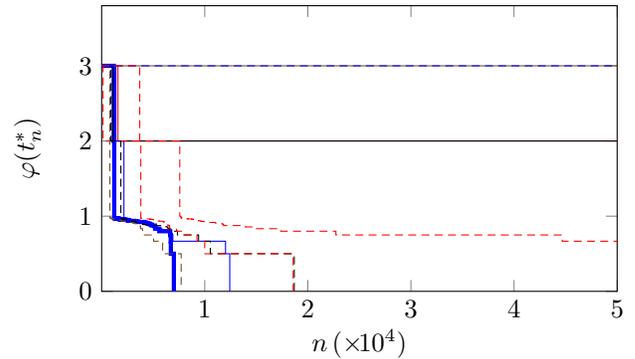

Figure~\ref{beagle:fig:expcasestudy} shows the evolution of the 10 workers running on our case study application \casestudyapp{}.
Each of the 10 lines represents the fitness $\varphi$ of the best individual at generation $n$, in symbols $t^*_n$, found by the corresponding worker.
A line hitting the $x$ axis corresponds to a worker that found a successful test $t$, i.e., such that $\varphi(t) = 0$.
Instead, a plot hitting the vertical line $x = 5 \times 10^4$ denotes a false negative result of the worker.

Out of 10 workers, 5 converged to a successful test.
All of them found the test in less than $20000$ generations.
The first worker to find a successful test (bold line) did it in $7006$ generations.
The resulting test was: 
click(9,225)$^-$, click(130,19), type(\texttt{"\textquotesingle <scr\textquotesingle Ipt\textquotesingle >\textquotesingle ale\textquotesingle rt\textquotesingle(9)</script>\textquotesingle"}), click(208,9)$^-$, click(41,79), click(24,13).
The first click event (labeled with $-$) is immaterial as it hits no control in the pages, i.e., it amounts to clicking on the background.
The second click sets the focus on the user name field.
Then, the type event injects the payload.
After another void click, the test terminates by clicking on submit and, then, on confirm.
Notice that the filter on the user name field removes the \textquotesingle{} from the payload which then results in \texttt{"<scrIpt>alert(9)</script>"}.

It is interesting to notice that one worker terminates with $\varphi(t^*) = 0.66$ (dashed, red line).
This means that the worker reached \texttt{welcome.php} but it could not find a payload for the injection.
In particular, the payload was \texttt{"<scripT>alert(0nyu:O)</\textquotesingle\textquotesingle\textquotesingle script>"}.
Although partial, the test generated by the worker may be of interest for a human analyst.
As a matter of fact, it provides a useful hint and a starting point for manually testing the AUT.

We also used our working example to evaluate the accuracy of \toolname{}. 
We executed \toolname{} 10 times on \casestudyapp{} (recall that each execution consists of 10 concurrent workers).
According to our experiments, the false negatives rate was 0\% for the executions, i.e., all the 10 runs of \toolname{} detected the vulnerability, and 56\% for the workers.

\subsection{WackoPicko benchmark}

To assess \toolname{} we applied it to the web application \emph{WackoPicko},
presented in~\cite{Vigna10johnny}.
WackoPicko contains 15 vulnerabilities and it has been used as a benchmark for both
white-box and black-box testing tools.
The complete list of the vulnerabilities disseminated in WackoPicko is given in Table~\ref{tab:wackopicko}.

\begin{table}[t]
\caption{The vulnerabilities of WackoPicko.}
\label{tab:wackopicko}
\begin{center}
\begin{tabular}{ @{\hspace{0pt}} r @{\hspace{3pt}} l c @{\hspace{20pt}} r @{\hspace{3pt}} l c @{\hspace{0pt}}}
\toprule
{\sc \#} & {\sc Description} & {\sc Test} & {\sc \#} & {\sc Description} & {\sc Test} \\
\toprule
1 & Reflected XSS & yes & 9 & Command-line injection & yes \\
2 & Stored XSS & yes & 10 & File inclusion & no$^\flat$ \\
3 & SessionID vulnerability & no$^\dagger$ & 11 & Parameter manipulation & no$^\dagger$ \\
4 & Stored SQL injection & yes & 12 & Reflected XSS behind JS & yes \\
5 & Reflected SQL injection & yes & 13 & Logic flaw & no$^\flat$ \\
6 & Directory traversal & no$^\flat$ & 14 & Reflected XSS behind Flash & yes \\
7 & Multi-step stored XSS & yes & 15 & Weak username/password & no$^\flat$ \\
8 & Forcefull browsing & no$^\flat$ \\ 
\bottomrule
\multicolumn{6}{@{\hspace{0pt}} l}{$\dagger$: extension needed $\qquad\flat$: out of scope} \\
\bottomrule
\end{tabular}
\end{center}
\end{table}

We refer the reader to~\cite{Vigna10johnny} for a detailed description of the vulnerabilities.
Five of the vulnerabilities in Table~\ref{tab:wackopicko} do not lay under our application conditions ($\flat$).
This is the case, for instance, of the directory traversal vulnerability (6) which allows an attacker to illegally access the file system.
This vulnerability is due to a misconfiguration of the execution platform and it does not occur in the AUT.
Two other vulnerabilities would require some major extensions of our prototype ($\dagger$).
For instance, the parameter manipulation vulnerability exploit requires to
directly modify URL parameters instead of performing user actions on the page
and \toolname{} is unable to simulate this behaviour with its set of actions.
The remaining 8 vulnerabilities form our benchmark.

\begin{figure*}
\includegraphics[width=\textwidth]{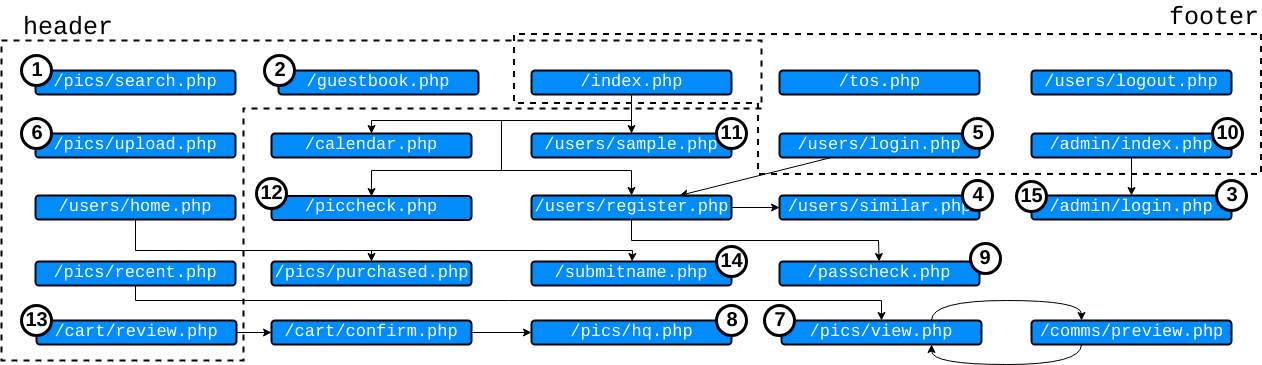}
\caption{An excerpt of the call graph of WackoPicko.}
\label{beagle:fig:wackopicko-cg}
\end{figure*}


WackoPicko consists of 2510 lines of PHP code grouped in 48 files, also including JavaScript and Adobe Flash components.
Figure~\ref{beagle:fig:wackopicko-cg} shows an excerpt of the call graph of the WackoPicko web application.
The entry point is the page \verb!/index.php!.
Each page includes a header and footer section (on top and bottom respectively).
Both the header and footer contain two menus with links to some pages (dashed boxes in Figure~\ref{beagle:fig:wackopicko-cg}).
These pages are thus accessible from any other page, i.e., each page in the call graph has an edge pointing to each page of the header and footer.
For the sake of readability we omit these edges.
The numbered, circular labels denote the presence of one of the vulnerabilities of Table~\ref{tab:wackopicko} in the page.

\begin{figure*}
\begin{center}
\input{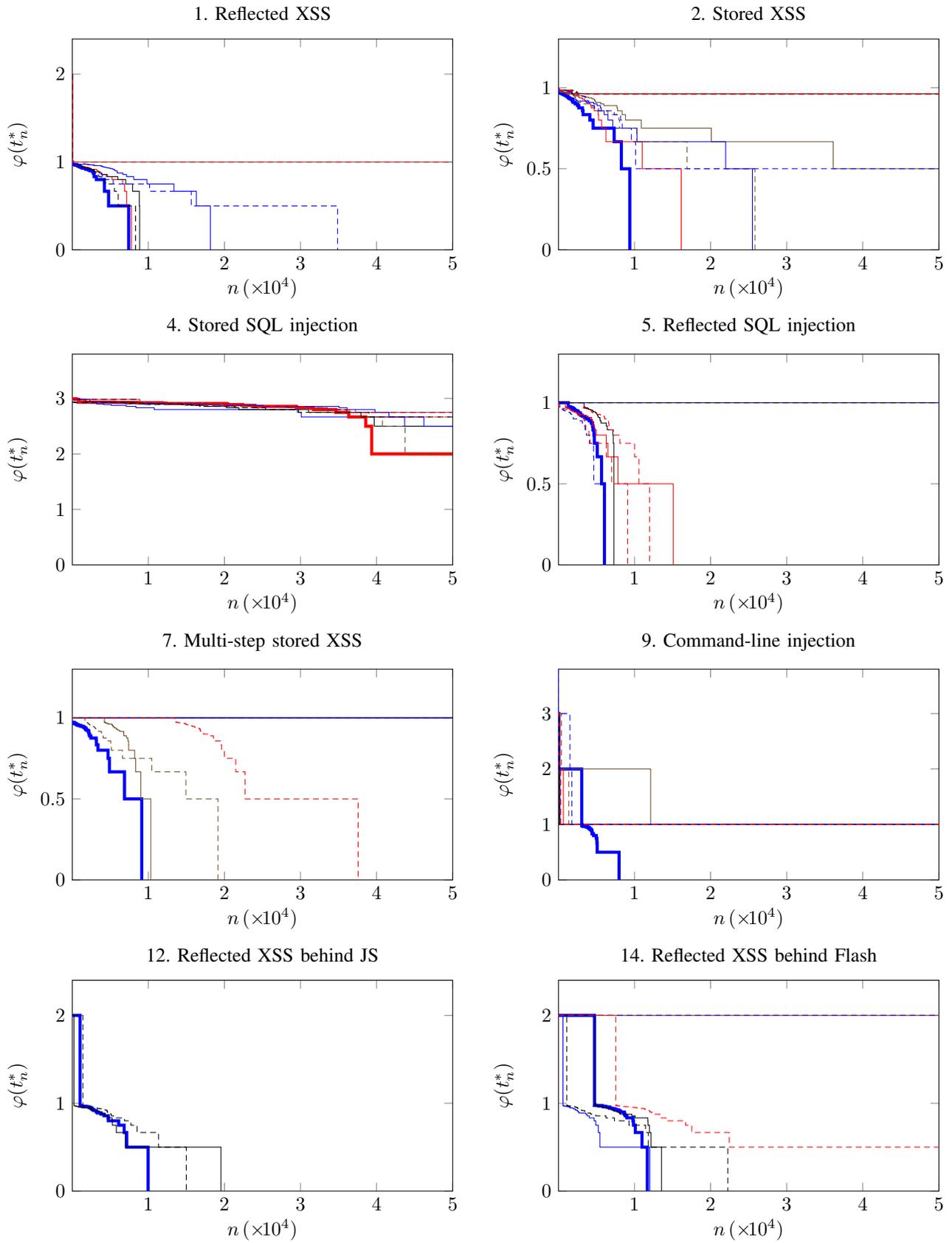}
\end{center}
\caption{Application to the WackoPicko benchmark.}
\label{beagle:fig:wackopickoexp}
\end{figure*}

\begin{table}
\caption{Payloads and performance of Beagle.}
\label{tab:wackopickoexp}
\begin{center}
\rowcolors{2}{lightgray!30}{white}
\begin{tabular}{r l r r r}
\toprule
\# & {\sc WSR} & {\sc Gen.} & {\sc Time} & {\sc Injected payload} \\ 
\toprule
1 & 6/10 & $7425$ & $787s$ & \texttt{<script>alert(44)</script>} \\
2 & 4/10 & $9380$ & $993s$ & \texttt{<script>alert(\textquotesingle`s\textquotesingle)</script>} \\
4 & 0/10 & $50000$ & $5132s$ & \texttt{Ac\textquotesingle qCM\#uYq6*4M-PaE} \\
5 & 6/10 & $6049$ & $640s$ & \texttt{admin\textquotesingle-- IgeBGMBL0`MnGUU99\#p}\\
7 & 4/10 & $9137$ & $968s$ & \texttt{n<script>alert(3)</script>FO} \\
9 & 1/10 & $7969$ & $844s$ & \texttt{q7d \&; ls \#X<\textquotesingle SI} \\
12 & 3/10 & $9970$ & $1056s$ & \texttt{<script>alert(3)</script>\textasciicircum JM} \\
14 & 4/10 & $11683$ & $1238s$ & \texttt{<script>alert(13)</sCRipt>} \\
\bottomrule
\end{tabular}
\end{center}
\end{table}

Figure~\ref{beagle:fig:wackopickoexp} shows the behavior of 10 \toolname{} workers for each of the considered vulnerabilities.
Again, the bold line indicates the best worker, i.e., the worker that detects the vulnerability for first.
The results of our experiments are reported in Table~\ref{tab:wackopickoexp}.
For each vulnerability we report the success rate of the workers (WSR), the number of generations and the execution of the best worker ({\sc Gen.} and {\sc Time}, respectively) and the generated injection ({\sc Injected payload}).
The injection is the malicious text typed by the generated test and submitted as a payload to the AUT. 

To better highlight the outcome of our experiments, consider the case of the multi-step stored XSS vulnerability.
A successful test has to browse through the section containing a list of recent pictures (\texttt{/pictures/recent.php}), click on one of them, inject the payload in the comment area in the picture page, click on the comment preview and then go back to the picture page.

\begin{figure*}
\begin{center}
\begin{tabular}{c @{\hspace{10pt}} c}
\includegraphics[width=0.45\textwidth]{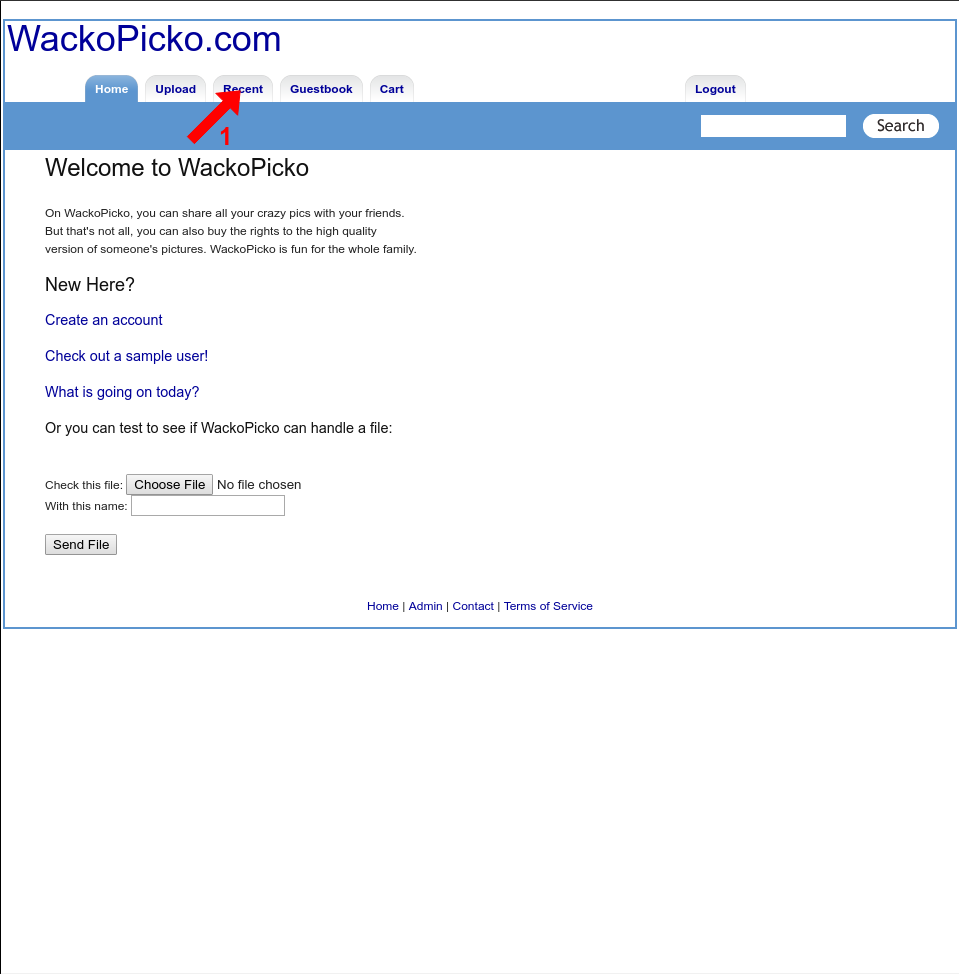} &
\includegraphics[width=0.45\textwidth]{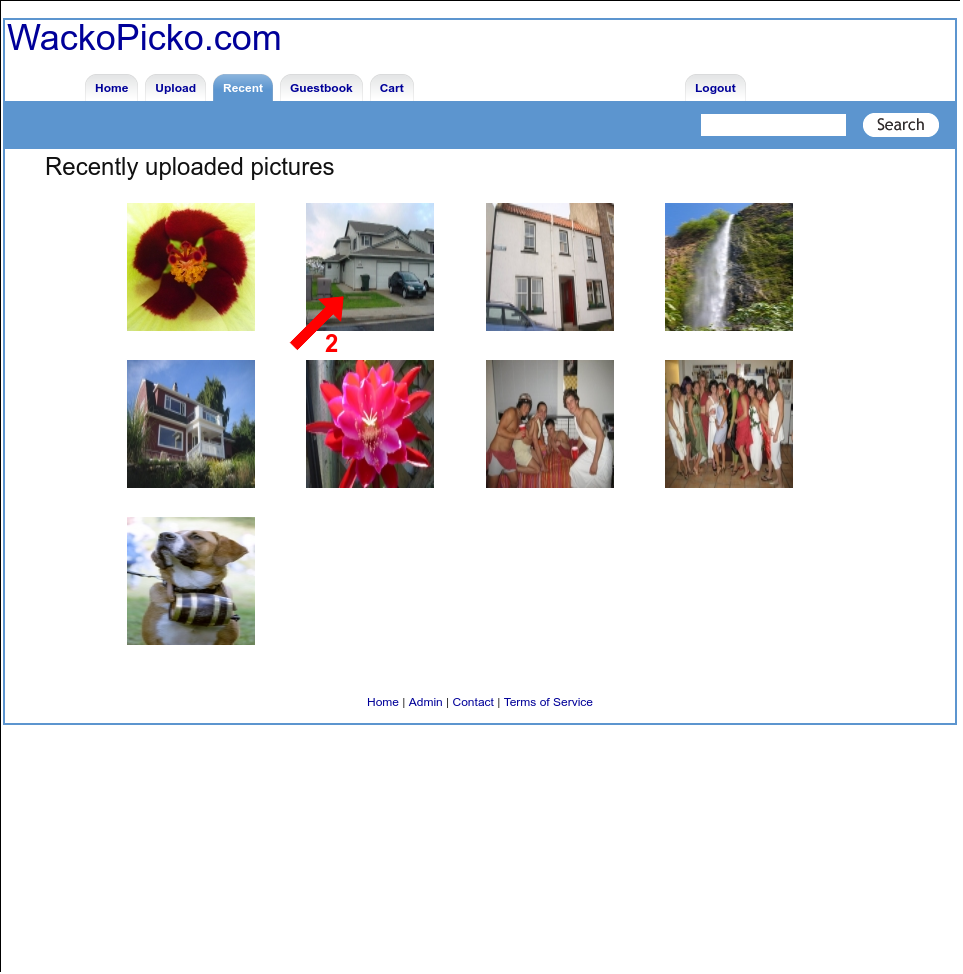} \\
\includegraphics[width=0.45\textwidth]{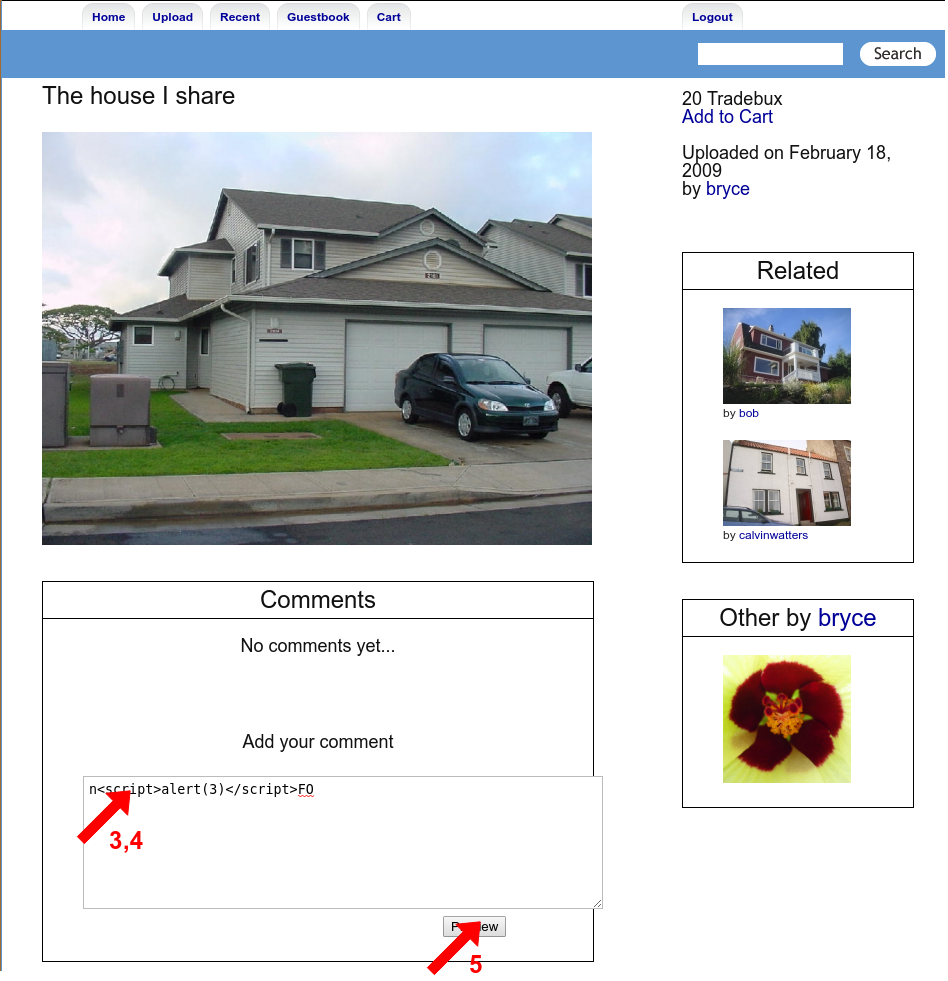} &
\includegraphics[width=0.45\textwidth]{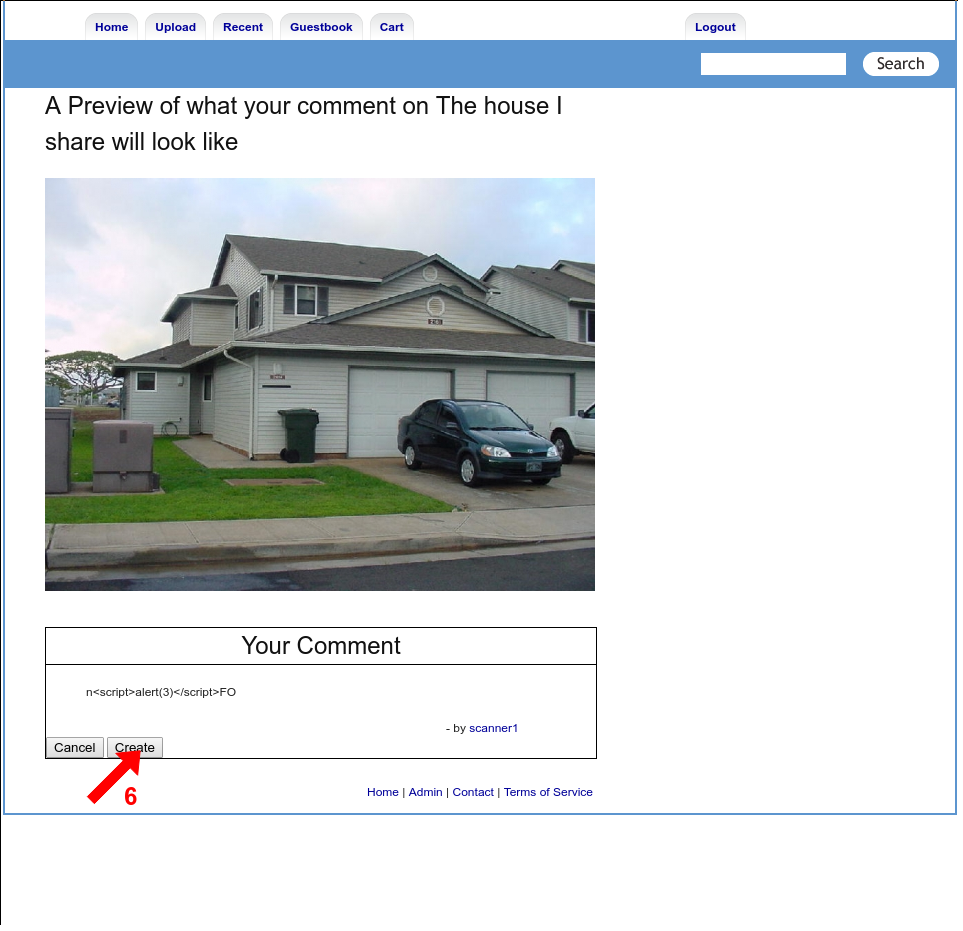} \\
\end{tabular}
\end{center}
\caption{The sequence of WackoPicko pages displayed during a successful test for the multi-step stored XSS vulnerability.}
\label{beagle:fig:spaccato}
\end{figure*} 

The sequence generated by \toolname{} appears in Figure~\ref{beagle:fig:spaccato} (red arrows indicate the positions of the events).
From top left to bottom right, the test proceeds as follow.
\begin{enumerate}
\item Click on the \emph{Recent} tab (top menu in the header);
\item click on the second picture of the first row;
\item click on the comment field;
\item type \texttt{"n<script>alert(3)</script>FO"};
\item click on the \emph{Preview} button;
\item click on the \emph{Create} button. 
\end{enumerate}
The last step causes the injected script to be executed.

\subsection{Comparison and discussion}

We ran three state-of-the-art web vulnerability scanners, i.e., OWASP ZAP, W3af and Vega, on WackoPicko.
Table~\ref{tab:toolscomparison} shows the vulnerabilities detected by each tool, including \toolname{}.
All the web scanners detected the Reflected XSS vulnerability (column 1) which can be exploited by directly submitting the injection payload to a single page of the AUT.
The same holds for the Reflected SQL injection vulnerability (column 5).
Finally, OWASP ZAP and Vega could also detect the Stored XSS (column 2).
As expected, all the web scanners failed in detecting the vulnerabilities that require to establish a non trivial interaction with the AUT.

\begin{table}
\caption{Comparison with other tools}
\label{tab:toolscomparison}
\begin{center}
\rowcolors{2}{lightgray!30}{white}
\begin{tabular}{l @{\hspace{10pt}} l l c c c c c c c c}
\toprule
{\sc Tool (configuration)} & {\sc 1} & {\sc 2} & {\sc 4} & {\sc 5} &
    {\sc 7} & {\sc 9} & {\sc 12} & {\sc 14} \\ 
\toprule
\textbf{Beagle}             &\checkmark{} &\checkmark{}& &\checkmark{}&\checkmark{}&\checkmark{}&\checkmark{}&\checkmark{} \\
\hline
OWASP ZAP (Standard mode)   &\checkmark{}&\checkmark{}& &\checkmark{}& & & & \\
OWASP ZAP (Attack mode)     &\checkmark{}&\checkmark{}& &\checkmark{}& & & & \\
W3af (fast-scan)            &\checkmark{}& & &\checkmark{}& & & & \\
W3af (full-audit)           &\checkmark{}& & &\checkmark{}& & & & \\
W3af (OWASP Top 10)         &\checkmark{}& & &\checkmark{}& & & & \\
Vega (Default)              &\checkmark{}&\checkmark{}& &\checkmark{}& & & & \\

\bottomrule
\end{tabular}
\end{center}
\end{table}

Moreover, we report the payloads generated by OWASP ZAP, W3af and Vega in
Table~\ref{tab:toolspayloads}.
Both OWASP ZAP and Vega rely on predefined dictionaries of injection payloads.
Such payloads are useful for a human analyst but suffers from few limitations.
Mainly, they cannot bypass payloads filters and blacklists.
On the other hand, W3af generates the payload through a pattern-based strategy.
W3af patterns are more flexible since they can adapt to multiple contexts.
Nevertheless, to be generic, patterns typically consist of complex rules that result in payloads that are harder to read for a human being.

\begin{table*}
\caption{Payloads injected by OWASP ZAP, Vega and {\sc W3af}.}
\label{tab:toolspayloads}
\begin{center}
\rowcolors{2}{lightgray!30}{white}
\begin{tabular}{r c c }
\toprule
\multicolumn{3}{c}{{\sc OWASP ZAP}} \\
\toprule
{\#} & {\sc Standard mode} & {\sc Attack mode} \\
\toprule
1 & \texttt{"><script>alert(1);</script>} & \texttt{"><script>alert(1);</script>} \\
2 & \texttt{</p><script>alert(1);</script><p>} & \texttt{</p><script>alert(1);</script><p>} \\
4 & & \\
5 & \texttt{ZAP' AND '1'='1' -- -}
  & \texttt{scanner1' AND '1'='1' -- } \\
7 & & \\
9 & & \\
12 & & \\
14 & & \\
\bottomrule
\end{tabular}
\hspace{88pt}
\rowcolors{2}{lightgray!30}{white}
\begin{tabular}{ l  c }
\toprule
\multicolumn{2}{c}{{\sc Vega}} \\
\toprule
{\#} & {\sc Default} \\
\toprule
1 & \texttt{1-->">'>"'} \\
2 & \texttt{vega' -->">'>"'} \\
4 & \\
5 & \texttt{Joey"'} \\
7 & \\
9 & \\
12 & \\
14 & \\
\bottomrule
\end{tabular}

\bigskip
\rowcolors{2}{lightgray!30}{white}
\begin{tabular}{r c c c }
\toprule
\multicolumn{4}{c}{{\sc W3af}} \\
\toprule
{\#} & {\sc Fast scan} & {\sc Full audit} & {\sc OWASP Top 10} \\
\toprule
1 & \begin{tabular}{@{} l @{}}
\rowcolor{lightgray!30} \texttt{v9ypk<v9ypkv9ypk-->v9ypkv9ypk*/v9ypk} \\
\rowcolor{lightgray!30} \texttt{v9ypk*/:("'v9ypkv9ypk:v9ypkv9ypk} \\
\rowcolor{lightgray!30} \texttt{v9ypkv9ypk"v9ypkv9ypk'v9ypkv9ypk`v9y} \\
\rowcolor{lightgray!30} \texttt{pkv9ypk =v9ypk}\end{tabular} & 
\begin{tabular}{@{} l @{}}
\rowcolor{lightgray!30} \texttt{coe5r<coe5rcoe5r-->coe5rcoe5r*/coe5r} \\
\rowcolor{lightgray!30} \texttt{coe5r*/:("'coe5rcoe5r:coe5rcoe5r} \\
\rowcolor{lightgray!30} \texttt{coe5rcoe5r"coe5rcoe5r'coe5rcoe5r`coe} \\
\rowcolor{lightgray!30} \texttt{5rcoe5r =coe5r}\end{tabular} &
\begin{tabular}{@{} l @{}}
\rowcolor{lightgray!30} \texttt{dmem4<dmem4dmem4-->dmem4dmem4*/dmem4} \\
\rowcolor{lightgray!30} \texttt{dmem4*/:("'dmem4dmem4:dmem4dmem4} \\ 
\rowcolor{lightgray!30} \texttt{dmem4dmem4"dmem4dmem4'dmem4dmem4`dme} \\
\rowcolor{lightgray!30} \texttt{m4dmem4 =dmem4}\end{tabular} \\
2 &  &  &  \\
4 &  &  &  \\
5 & \texttt{a'b"c'd"} & \texttt{a'b"c'd"} & \texttt{a'b"c'd"} \\
7 &  &  &  \\
9 &  &  &  \\
12 &  &  &  \\
14 &  &  &  \\
\bottomrule 
\end{tabular}
\end{center}
\end{table*}


\section{Related work}
\label{beagle:sec:related}

The problem of testing event-based systems to discover their vulnerabilities has received significant attention in the last decades and many proposals have been put forward.
Several proposals target some specific GUI environments (e.g., Android) and a direct comparison with our approach is not straightforward.
Nevertheless, in the following, we discuss the similarity and differences w.r.t. our proposal.

\noindent
\textbf{Search-based testing.}
According to~\cite{McMinn11search}, search-based testing includes the testing methodologies that automate the test generation, execution and evaluation through some heuristics.

Our approach belongs to this category and, in particular, to the \emph{genetic testing} techniques.
In this case, the search heuristic is defined in terms of a fitness function computed over a set of tests, namely the \emph{population}.
A typical fitness function is the AUT code coverage, i.e., the test that leads to the execution of the largest portion of the AUT code receives the highest score.
This is the approach followed, for instance, in~\cite{budd81mutation,pargas99gentest,michael01evogentest,bueno02automaticgentest}. 
At the best of our knowledge, this is the first proposal for a genetic testing technique for an aimed vulnerability testing generation.

Several tools implement the genetic technique for the generation of coverage tests. 
\emph{EvoDroid}~\cite{Mahmood14evodroid} is an evolutionary GUI-based, coverage testing tool for Android applications.
Its selection process favors the tests that maximize the coverage of the AUT GUI elements.
Unlike \toolname{}, EvoDroid cannot converge toward the execution of a specific vulnerability.
The reason is that, although its fitness function can be compared to our call distance function, EvoDroid does not have a concept of contract distance.
As we discussed above, the contract distance is crucial for the aimed search strategy of \toolname{}.

Dynodroid~\cite{machiry13dynodroid} is a system for automatically generating relevant inputs for an Android AUT.
Interestingly, Dynodroid does not modify the running application, but it heavily instruments the Android framework to get feedback directly from the event handlers.
Dynodroid binds each user action, e.g., tapping at certain coordinates on the screen, to a corresponding GUI event, e.g., a button click.
Then, it identifies the least stimulated GUI elements and generates new tests to increase the interaction with them.
This approach is called \emph{frequency strategy}.
Although it can effectively lead to a good coverage of the AUT behavior, Dynodroid cannot identify the critical inputs that force the AUT to run a vulnerable instruction.

GUI Ripper~\cite{amalfitano12toolset,amalfitano12using}, systematically generates GUI events and fires them on the AUT.
Starting from an initial state, it explores the AUT using a depth-first search strategy.
For each state, it keeps track of all the possible actions.
At each execution, it resets the AUT and tries to perform actions never attempted before in the current state to extend the AUT coverage.

\emph{Random testing}~\cite{breuer71random} is a common form of search-based testing that works by sending random GUI events, e.g., clicks and keystrokes, to the AUT.
Often, the modern development frameworks natively include a random testing engine, e.g., the Android \textit{monkey} tool.\footnote{https://developer.android.com/studio/test/monkey.html}
As we showed in our working example, random tasting has a very remote probability of spotting out a vulnerability.

The MAVeriC~\cite{armando14mavericstatic,Armando15mavericdynamic} platform integrates a static and analysis environment for Android applications.
The dynamic analysis consists of a TEE that randomly stimulates the AUT in order to trigger some illegal behavior.
Nevertheless, the model inference part only aims at providing a human analyst with extra knowledge about the AUT, and no automatic test case generation is supported.

\noindent
\textbf{Model-based testing.}
The idea behind model-based testing~\cite{Fraser2009TMC} is to take advantage of a model of the AUT to drive the test case generation or execution.
For instance, a model checker can be exploited to statically verify whether a vulnerability is present in a model of the AUT.
If so, the model checker returns a counterexample that can be tested against the AUT to confirm the vulnerability.
However, in most cases, a proper model of the AUT is not available.
Moreover, when the model does not precisely represent the behavior of the AUT, the risk of false positives and false negatives grows.

QED~\cite{martin08qed} is a goal-directed model-checking system that automatically generates attacks exploiting taint-based vulnerabilities in Java web applications.
Vernotte et al.~\cite{vernotte14msxss} propose a \emph{pattern-driven model-based} approach (PMVT) to detect a multi-step XSS vulnerability in a codebase.
The approach relies on software called CertifyIt~\cite{legeard13certifyit} to generate tests for the AUT.
CertifyIt takes a UML subset as an input to infer the behavior of a web application.
Using a UML model has some drawbacks since there is no guarantee that the code behaves as its specification says.
This is particularly true for the vulnerabilities that typically do not appear in the specification.

\noindent
\textbf{Symbolic and hybrid testing.}
The idea behind this approach is to replace the standard, concrete semantics of a programming language with an abstract, symbolic one.
Naively, the symbolic semantics replaces the actual values of the variables (e.g., $x \mapsfrom 5$) with a predicate (e.g., $x \geq 2 \wedge x \neq 7$).
In this way, \emph{symbolic execution}~\cite{King76symbolic} permits to replace a large number of tests with a single, symbolic one.
The main limitation of this approach is its poor scalability on large, real-life software.
Hybrid testing techniques combine two or more approaches to improve the TEE regarding scalability or effectiveness.
For instance, \emph{concolic testing}~\cite{sen07concolic} is a variant of symbolic testing where the application is executed alternating concrete and symbolic values.
Symbolic constraints are used to generate valid concrete testing values incrementally.
\emph{Hybrid concolic testing}~\cite{hybridconcolic} merges random and concolic testing in order to improve the scalability of the concolic testing.
One of the main issues with the symbolic approach is that it requires the target language to be provided with a formal semantics.
Most programming languages do not have it.
Our approach can be applied without this limitation and, in the presence of formal semantics, it can benefit from it for the automated contract inference.

\noindent
\textbf{Taint-based analysis.}
A taint analysis tool inspects the code of the AUT to find the dependencies between its variables and to understand how they influence each other.

FlowDroid~\cite{arzt2014flowdroid} is a taint analysis tool that statically checks the application code.
In particular, it analyses the data flow logic of the application to detect whether data can move from a source, e.g., an input field, to a destination, e.g., a SQL query.
In doing that, FlowDroid introduces approximations that may cause false positive/negatives.
Thus, the illegal data flow must be confirmed through a test case.

Similarly, ARDILLA~\cite{Kiezun09ardilla} generates initial test inputs for SQL injection and XSS vulnerabilities by symbolically tracking tainted values through the execution.
It then mutates the input to produce actual exploits according to the information gathered during the taint analysis phase.

These techniques can effectively discover an injection flaw vulnerability.
However, they typically do not apply to the event-based systems.
As a matter of fact, the events may influence the execution flow of the program that needs to be modeled, e.g., by introducing approximations that increase the risk of false results.

RIPS~\cite{dahse2014simulation} evaluates the built-in PHP sanitization functions and, in particular, it checks whether an execution flow is tainted or not.
Chainsaw~\cite{alhuzali2016chainsaw} uses pure static analysis to build concrete exploits in multi-tiered web applications.
NAVEX~\cite{alhuzali2018navex} is an extension of Chainsaw that identifies vulnerabilities as graph queries on a Navigation Graph.
It marks vulnerable function calls as sinks and it tries to traverse back to the source of its arguments, passing through the sanitization functions.
NAVEX relies on a predefined \emph{attack dictionary} and cannot generate new attack payloads on-the-fly.

\section{Conclusion}
\label{beagle:sec:conclusion}

In this paper we introduced a new approach for the automatic vulnerability testing of event-based systems.
Our proposal uses a collaborative, co-evolutionary algorithm that relies on a novel contract distance function.
We applied our technique in the context of web applications vulnerability testing and we implemented a working prototype called \toolname{}.

We applied \toolname{} to a case study application suffering from a multi-step stored XSS vulnerability.
Such vulnerability is well known to be a very challenging one for automatic vulnerability scanners.
Moreover, we carried out an experimentation on a benchmark of vulnerabilities, WackoPicko.
Our results show that \toolname{} effectively discovered all the vulnerabilities but one,
whereas other mainstream web scanners failed at discovering and exploiting most of the vulnerabilities.

We plan to apply \toolname{} to real world web applications in order to better evaluate its scalability and performances. 
Moreover, we will extend our approach to other programming frameworks and languages.
All these directions account as future work.


\paragraph*{Acknowledgment}
This research was conducted when Andrea Valenza was at University of Genova.

\bibliographystyle{plain}
\bibliography{biblio}

\begin{thebibliography}{10}

\bibitem{alhuzali2016chainsaw}
Abeer Alhuzali, Birhanu Eshete, Rigel Gjomemo, and VN~Venkatakrishnan.
\newblock Chainsaw: Chained automated workflow-based exploit generation.
\newblock In {\em Proceedings of the 2016 ACM SIGSAC Conference on Computer and
  Communications Security}, pages 641--652. ACM, 2016.

\bibitem{alhuzali2018navex}
Abeer Alhuzali, Rigel Gjomemo, Birhanu Eshete, and VN~Venkatakrishnan.
\newblock Navex: precise and scalable exploit generation for dynamic web
  applications.
\newblock In {\em 27th USENIX Security Symposium (USENIX Security '18)}, pages
  377--392. USENIX Association, 2018.

\bibitem{amalfitano12toolset}
Domenico Amalfitano, Anna~Rita Fasolino, Porfirio Tramontana, Salvatore
  De~Carmine, and Gennaro Imparato.
\newblock A toolset for gui testing of android applications.
\newblock In {\em Software Maintenance (ICSM), 2012 28th IEEE International
  Conference on}, pages 650--653. IEEE, 2012.

\bibitem{amalfitano12using}
Domenico Amalfitano, Anna~Rita Fasolino, Porfirio Tramontana, Salvatore
  De~Carmine, and Atif~M Memon.
\newblock Using gui ripping for automated testing of android applications.
\newblock In {\em Proceedings of the 27th IEEE/ACM International Conference on
  Automated Software Engineering}, pages 258--261. ACM, 2012.

\bibitem{Armando15mavericdynamic}
Alessandro Armando, Gianluca Bocci, Gabriele Costa, Rocco Mammoliti, Alessio
  Merlo, Silvio Ranise, Riccarto Traverso, and Andrea Valenza.
\newblock Mobile app security assessment with the maveric dynamic analysis
  module.
\newblock In {\em Proceedings of the 7th ACM CCS International Workshop on
  Managing Insider Security Threats}, MIST '15, pages 41--49, New York, NY,
  USA, 2015. ACM.

\bibitem{armando14mavericstatic}
Alessandro Armando, Giantonio Chiarelli, Gabriele Costa, Gabriele De~Maglie,
  Rocco Mammoliti, and Alessio Merlo.
\newblock Mobile app security analysis with the maveric static analysis module.
\newblock {\em JoWUA}, 5(4):103--119, 2014.

\bibitem{arzt2014flowdroid}
Steven Arzt, Siegfried Rasthofer, Christian Fritz, Eric Bodden, Alexandre
  Bartel, Jacques Klein, Yves Le~Traon, Damien Octeau, and Patrick McDaniel.
\newblock Flowdroid: Precise context, flow, field, object-sensitive and
  lifecycle-aware taint analysis for android apps.
\newblock {\em Acm Sigplan Notices}, 49(6):259--269, 2014.

\bibitem{wpcal}
Marcello~M. Bonsangue and Joost~N. Kok.
\newblock {The Weakest Precondition Calculus: Recursion and Duality}.
\newblock {\em Formal Aspects of Computing}, 6(1):788--800, November 1994.

\bibitem{breuer71random}
Melvin~A Breuer.
\newblock A random and an algorithmic technique for fault detection test
  generation for sequential circuits.
\newblock {\em IEEE Transactions on Computers}, 100(11):1364--1370, 1971.

\bibitem{budd81mutation}
Timothy~A Budd.
\newblock Mutation analysis: Ideas, examples, problems and prospects.
\newblock {\em Computer Program Testing}, 8:i29--l48, 1981.

\bibitem{bueno02automaticgentest}
Paulo Marcos~Siqueira Bueno and Mario Jino.
\newblock Automatic test data generation for program paths using genetic
  algorithms.
\newblock {\em International Journal of Software Engineering and Knowledge
  Engineering}, 12(06):691--709, 2002.

\bibitem{dahse2014simulation}
Johannes Dahse and Thorsten Holz.
\newblock Simulation of built-in php features for precise static code analysis.
\newblock In {\em NDSS}. Citeseer, 2014.

\bibitem{smt}
Leonardo De~Moura and Nikolaj Bj{\o}rner.
\newblock Satisfiability modulo theories: Introduction and applications.
\newblock {\em Commun. ACM}, 54(9):69--77, September 2011.

\bibitem{Vigna10johnny}
Adam Doup{\'e}, Marco Cova, and Giovanni Vigna.
\newblock Why johnny can't pentest: An analysis of black-box web vulnerability
  scanners.
\newblock In {\em Proceedings of the 7th International Conference on Detection
  of Intrusions and Malware, and Vulnerability Assessment}, DIMVA'10, pages
  111--131, Berlin, Heidelberg, 2010. Springer-Verlag.

\bibitem{DEAP2012}
F\'elix-Antoine Fortin, Fran\c{c}ois-Michel {De Rainville}, Marc-Andr\'e
  Gardner, Marc Parizeau, and Christian Gagn\'e.
\newblock {DEAP}: Evolutionary algorithms made easy.
\newblock {\em Journal of Machine Learning Research}, 13:2171--2175, jul 2012.

\bibitem{Fraser2009TMC}
Gordon Fraser, Franz Wotawa, and Paul~E. Ammann.
\newblock Testing with model checkers: A survey.
\newblock {\em Softw. Test. Verif. Reliab.}, 19(3):215--261, September 2009.

\bibitem{Kiezun09ardilla}
Adam Kie{\.z}un, Philip~J. Guo, Karthick Jayaraman, and Michael~D. Ernst.
\newblock Automatic creation of {SQL} injection and cross-site scripting
  attacks.
\newblock In {\em ICSE 2009, Proceedings of the 31st International Conference
  on Software Engineering}, pages 199--209, Vancouver, BC, Canada, May 2009.

\bibitem{King76symbolic}
James~C. King.
\newblock Symbolic execution and program testing.
\newblock {\em Commun. ACM}, 19(7):385--394, July 1976.

\bibitem{legeard13certifyit}
Bruno Legeard and Arnaud Bouzy.
\newblock Smartesting certifyit: Model-based testing for enterprise it.
\newblock In {\em Software Testing, Verification and Validation (ICST), 2013
  IEEE Sixth International Conference on}, pages 391--397. IEEE, 2013.

\bibitem{levdist}
VI~Levenshtein.
\newblock {Binary Codes Capable of Correcting Deletions, Insertions and
  Reversals}.
\newblock {\em Soviet Physics Doklady}, 10:707, 1966.

\bibitem{machiry13dynodroid}
Aravind Machiry, Rohan Tahiliani, and Mayur Naik.
\newblock Dynodroid: An input generation system for android apps.
\newblock In {\em Proceedings of the 2013 9th Joint Meeting on Foundations of
  Software Engineering}, pages 224--234. ACM, 2013.

\bibitem{Mahmood14evodroid}
Riyadh Mahmood, Nariman Mirzaei, and Sam Malek.
\newblock Evodroid: Segmented evolutionary testing of android apps.
\newblock In {\em Proceedings of the 22Nd ACM SIGSOFT International Symposium
  on Foundations of Software Engineering}, FSE 2014, pages 599--609, New York,
  NY, USA, 2014. ACM.

\bibitem{hybridconcolic}
Rupak Majumdar and Koushik Sen.
\newblock {Hybrid Concolic Testing}.
\newblock In {\em Proceedings of the 29th International Conference on Software
  Engineering}, ICSE '07, pages 416--426, Washington, DC, USA, 2007. IEEE
  Computer Society.

\bibitem{martin08qed}
Michael Martin and Monica~S Lam.
\newblock Automatic generation of xss and sql injection attacks with
  goal-directed model checking.
\newblock In {\em Proceedings of the 17th conference on Security symposium},
  pages 31--43. USENIX Association, 2008.

\bibitem{McMinn11search}
Phil McMinn.
\newblock {Search-Based Software Testing: Past, Present and Future}.
\newblock In {\em Proceedings of the 2011 IEEE Fourth International Conference
  on Software Testing, Verification and Validation Workshops}, ICSTW '11, pages
  153--163, Washington, DC, USA, 2011. IEEE Computer Society.

\bibitem{meyercdd}
Bertrand Meyer.
\newblock {Contract-Driven Development}.
\newblock In Matthew~B. Dwyer and Ant{\'o}nia Lopes, editors, {\em Fundamental
  Approaches to Software Engineering}, pages 11--11, Berlin, Heidelberg, 2007.
  Springer Berlin Heidelberg.

\bibitem{michael01evogentest}
Christoph~C. Michael, Gary McGraw, and Michael~A Schatz.
\newblock Generating software test data by evolution.
\newblock {\em IEEE transactions on software engineering}, 27(12):1085--1110,
  2001.

\bibitem{Papadimitriou81integer}
Christos~H. Papadimitriou.
\newblock {On the Complexity of Integer Programming}.
\newblock {\em Journal of the ACM}, 28(4):765--768, October 1981.

\bibitem{pargas99gentest}
Roy~P Pargas, Mary~Jean Harrold, and Robert~R Peck.
\newblock Test-data generation using genetic algorithms.
\newblock {\em Software Testing Verification and Reliability}, 9(4):263--282,
  1999.

\bibitem{OWASPtg}
Open Web Application~Security Project.
\newblock {OWASP Testing Guide 4.0}.
\newblock Available at \url{https://www.owasp.org/images/1/19/OTGv4.pdf}.

\bibitem{sen07concolic}
Koushik Sen.
\newblock Concolic testing.
\newblock In {\em Proceedings of the twenty-second IEEE/ACM international
  conference on Automated software engineering}, pages 571--572. ACM, 2007.

\bibitem{vernotte14msxss}
Alexandre Vernotte, Fr{\'e}d{\'e}ric Dadeau, Franck Lebeau, Bruno Legeard,
  Fabien Peureux, and Fran{\c{c}}ois Piat.
\newblock Efficient detection of multi-step cross-site scripting
  vulnerabilities.
\newblock In {\em International Conference on Information Systems Security},
  pages 358--377. Springer, 2014.

\bibitem{z3str}
Yunhui Zheng, Xiangyu Zhang, and Vijay Ganesh.
\newblock {Z3-str: A Z3-based String Solver for Web Application Analysis}.
\newblock In {\em Proceedings of the 2013 9th Joint Meeting on Foundations of
  Software Engineering}, ESEC/FSE 2013, pages 114--124, New York, NY, USA,
  2013. ACM.

\end{thebibliography}

\end{document}